\documentclass{aa}  
\usepackage{graphicx}
\usepackage{color}
\usepackage{graphicx}
\usepackage{lscape}
\usepackage{natbib}
\usepackage[varg]{txfonts}
\usepackage{url}
\usepackage{xspace}
\bibpunct{(}{)}{;}{a}{}{,}
\newcounter{Rco}
\newcommand{\Ionst}[1]{\setcounter{Rco}{#1}\Roman{Rco}}

\newcommand{\Ionw}[3]{\mbox{#1\,{\scriptsize\Ionst{#2}}~$\lambda\,#3$\,\AA}}

\newcommand{\Ionww}[3]{\mbox{#1\,{\scriptsize\Ionst{#2}}~$\lambda\lambda\,#3$\,\AA}}

\newcommand{\logg}{\mbox{$\log g$}\xspace}
\newcommand{\loggw}[1]{\mbox{$\log g\hspace{-0.5mm} =\hspace{-0.5mm}  #1$}}

\newcommand{\sga}{\raisebox{-0.10em}{$\stackrel{>}{{\mbox{\tiny $\sim$}}}$}}

\newcommand{\sla}{\raisebox{-0.10em}{$\stackrel{<}{{\mbox{\tiny $\sim$}}}$}}

\newcommand{\ta}[1]{\mbox{Tab.\,\ref{#1}}}
\newcommand{\sT}[1]{\mbox{(Tab.\,\ref{#1})}}
\newcommand{\Teff}{\mbox{$T_\mathrm{eff}$}\xspace}
\newcommand{\Teffw}[1]{\mbox{$\Teff\hspace{-0.5mm} =\hspace{-0.5mm} #1 \,\mathrm{K}$}}
\newcommand{\ebv}{\mbox{$E_{B-V}$}}

\newcommand{\deh}{\mbox{$N_\ion{D}{i}$}}
\newcommand{\nh}{\mbox{$N_\ion{H}{i}$}}

\newcommand{\Msol}{$M_\odot$}

\newcommand{\mmspr}{\hbox{}\hspace{+0.8cm}}
\newcommand{\smspr}{\hbox{}\hspace{+2.5mm}}
\newcommand{\smspl}{\hbox{}\hspace{+0.1mm}}

\newcommand{\gb}{\object{G191$-$B2B}\xspace}
\begin{document}
\title{The virtual observatory service TheoSSA: \\
       Establishing a database of synthetic stellar flux standards
      }
\subtitle{I. NLTE spectral analysis of the DA-type white dwarf \gb
           \thanks
           {Based on observations with the NASA/ESA Hubble Space Telescope, obtained at the Space Telescope Science 
            Institute, which is operated by the Association of Universities for Research in Astronomy, Inc., under 
            NASA contract NAS5-26666.
           }
           \thanks
           {Based on observations made with the NASA-CNES-CSA Far Ultraviolet Spectroscopic Explorer.
           }
           \thanks
           {Figures A.1 and A.2 are available in electronic form at
            http://www.aanda.org  (they are also available at the CDS in FITS format).
           }       
         }

\author{T\@. Rauch\inst{1} 
        \and 
        K\@. Werner\inst{1}
        \and 
        R\@. Bohlin\inst{2}
        \and 
        J\@. W\@. Kruk\inst{3}
        }
 
\institute{Institute for Astronomy and Astrophysics,
           Kepler Center for Astro and Particle Physics,
           Eberhard Karls University, 
           Sand 1,
           72076 T\"ubingen, 
           Germany,\\
           \email{rauch@astro.uni-tuebingen.de}
           \and
           Space Telescope Science Institute, 3700 San Martin Drive, Baltimore, MD \,21218, USA
           \and       
           NASA Goddard Space Flight Center, Greenbelt, MD\,20771, USA}

\date{Received 22 July 2013 / Accepted 22 August 2013}

\titlerunning{TheoSSA: Establishing a database of synthetic stellar flux standards. I. \gb}

\abstract {
           Hydrogen-rich, DA-type white dwarfs are particularly suited 
           as primary standard stars for flux calibration.
           State-of-the-art NLTE models
           consider opacities of species up to trans-iron elements and
           provide reliable synthetic stellar-atmosphere spectra to
           compare with observation.
          }
          {
           We will establish a database of theoretical spectra of
           stellar flux standards that are easily accessible 
           via a web interface. 
          }
          {
           In the framework of the Virtual Observatory,
           the German Astrophysical Virtual Observatory
           developed the registered service
           TheoSSA.
           It provides easy access to stellar 
           spectral energy distributions (SEDs) and is
           intended to ingest SEDs calculated by any model-atmosphere
           code.
           In case of the DA white dwarf \gb, we demonstrate that the
           model reproduces not only its overall continuum shape but
           also the numerous metal lines exhibited in its
           ultraviolet spectrum. 
          }
          {
           TheoSSA is in operation and contains presently a
           variety of SEDs for DA-type white dwarfs. It will be extended
           in the near future and can host SEDs of all primary and
           secondary flux standards. 
           The spectral analysis of \gb has
           shown that our hydrostatic models reproduce the observations best at 
           \Teffw{60\,000 \pm 2000} and 
           \loggw{7.60 \pm 0.05}. 
           We newly identified 
           \ion{Fe}{vi}, 
           \ion{Ni}{vi}, and
           \ion{Zn}{iv} lines.
           For the first time, we determined the
           photospheric zinc abundance with a
           logarithmic mass fraction of $-4.89$ 
           (7.5  $\times$ solar).
           The abundances of He (upper limit), C, N, O, Al, Si, O, P, S, Fe, Ni, Ge, and Sn 
           were precisely determined. Upper abundance limits of about 10\,\% solar 
           were derived for Ti, Cr, Mn, and Co.
          }
          {
           The TheoSSA database of theoretical SEDs of stellar flux standards 
           guarantees that the flux calibration of all astronomical data 
           and cross-calibration between different instruments can be based on the 
           same models and SEDs calculated with different model-atmosphere codes
           and are easy to compare. 
          }
         
\keywords{Standards --
          Stars: abundances -- 
          Stars: atmospheres -- 
          Stars: individual: \gb\ --
          Stars: white dwarfs --
          Virtual observatory tools
         }

\maketitle

\section{Introduction}
\label{sect:intro}

In the framework of the Virtual Observatory (VO), the 
German Astrophysical Virtual Observatory (GAVO) project provides 
synthetic stellar spectra on demand via the registered
Theoretical Stellar Spectra Access (TheoSSA) VO service 
\citep{rauch2008a,rauchnickelt2009,rauchetal2009}. 
These SEDs can be used for spectral analyses
\citep{rauchetal2010a, ringatrauch2010,rauchringat2011,ringatrauchwerner2012}
or serve as ionizing spectra for e.g\@. photoionization models
of ionized nebulae. The registered 
TMAW VO tool\footnote{T\"ubingen Model-Atmosphere WWW Interface, \url{http://astro.uni-tuebingen.de/~TMAW}}, 
that allows to calculate individual NLTE model atmospheres considering opacities of
H, He, C, N, O, Ne, Na, and Mg, provides additional SEDs which are
automatically ingested by TheoSSA. Figure\,\ref{fig:theossa} shows the
complete action scheme for a VO user to retrieve an SED.

\onlfig{
\begin{figure*}[ht!]
\begin{picture}(17.0,17.7)
\thicklines
%
%
%
\put( 4.5,17.0){\oval( 1.0, 1.0)[tl]}
\put(10.5,17.0){\oval( 1.0, 1.0)[tr]}
\put( 4.5,15.5){\oval( 1.0, 1.0)[bl]}
\put(10.5,15.5){\oval( 1.0, 1.0)[br]}
\put( 4.5,17.5){\line( 1, 0){ 6.0}}
\put( 4.5,15.0){\line( 1, 0){ 6.0}}
\put( 4.0,15.5){\line( 0, 1){ 1.5}}
\put(11.0,15.5){\line( 0, 1){ 1.5}}
\put( 4.0,16.5){\makebox( 7.0, 1.0)[c]{
{{\color{green}\sc TheoSSA} request via http://dc.g-vo.org/theossa:}
}}
\put( 4.0,16.0){\makebox( 7.0, 1.0)[c]{
{\color{red}$T_{\mathrm{eff}}$, $\log g$, $\{X_{\mathrm{i}}\}$}
}}
\put( 4.0,15.5){\makebox( 7.0, 1.0)[c]{
{standard flux table}
}}
\put( 4.0,15.0){\makebox( 7.0, 1.0)[c]{
{individual flux table: $\lambda$ interval, resolution}
}}
\put( 7.5,15.0){\vector( 0,-1){ 1.0}}
\put( 5.0,13.5){\oval( 1.0, 1.0)[tl]}
\put(10.0,13.5){\oval( 1.0, 1.0)[tr]}
\put( 5.0,12.5){\oval( 1.0, 1.0)[bl]}
\put(10.0,12.5){\oval( 1.0, 1.0)[br]}
\put( 5.0,14.0){\line( 1, 0){ 5.0}}
\put( 5.0,12.0){\line( 1, 0){ 5.0}}
\put( 4.5,12.5){\line( 0, 1){ 1.0}}
\put(10.5,12.5){\line( 0, 1){ 1.0}}
\put( 4.0,12.0){\makebox( 7.0, 3.0)[c]{
check of GAVO database:
}}
\put( 4.0,11.5){\makebox( 7.0, 3.0)[c]{
are requested parameters matched
}}
\put( 4.0,11.0){\makebox( 7.0, 3.0)[c]{
within tolerance limits?
}}
\put(10.5,13.0){\vector( 1, 0){ 3.0}}
\put(13.5,13.0){\vector(-1, 0){ 3.0}}
\put( 7.50,12.00){\line( 0,-1){ 0.5}}
\put( 4.75,11.50){\line( 1, 0){ 5.5}}
\put( 4.75,11.50){\vector( 0,-1){ 0.5}}
\put(10.25,11.50){\vector( 0,-1){ 0.5}}
%
%
%
\put( 3.0,10.5){\oval( 1.0, 1.0)[tl]}
\put( 6.5,10.5){\oval( 1.0, 1.0)[tr]}
\put( 3.0,10.0){\oval( 1.0, 1.0)[bl]}
\put( 6.5,10.0){\oval( 1.0, 1.0)[br]}
\put( 3.0,11.0){\line( 1, 0){ 3.5}}
\put( 3.0, 9.5){\line( 1, 0){ 3.5}}
\put( 2.5,10.0){\line( 0, 1){ 0.5}}
\put( 7.0,10.0){\line( 0, 1){ 0.5}}
\put( 2.0,10.0){\makebox( 5.5, 1.0)[c]{
{\color{blue}yes}:
}}
\put( 2.0, 9.5){\makebox( 5.5, 1.0)[c]{
offer existing model
}}
%
%
%
\put( 8.5,10.5){\oval( 1.0, 1.0)[tl]}
\put(12.0,10.5){\oval( 1.0, 1.0)[tr]}
\put( 8.5,10.0){\oval( 1.0, 1.0)[bl]}
\put(12.0,10.0){\oval( 1.0, 1.0)[br]}
\put( 8.5,11.0){\line( 1, 0){ 3.5}}
\put( 8.5, 9.5){\line( 1, 0){ 3.5}}
\put( 8.0,10.0){\line( 0, 1){ 0.5}}
\put(12.5,10.0){\line( 0, 1){ 0.5}}
\put( 7.5,10.0){\makebox( 5.5, 1.0)[c]{
{\color{red}no}:
}}
\put( 7.5, 9.50){\makebox( 5.5, 1.0)[c]{
calculate new {\color{green}\sc TMAW} model
}}
\put(12.5,10.25){\vector( 1, 0){ 1.0}}
\put( 4.75, 9.50){\line( 0,-1){ 0.5}}
\put( 2.75, 9.00){\line( 1, 0){ 4.0}}
\put( 2.75, 9.00){\vector( 0,-1){ 0.5}}
\put( 6.75, 9.00){\vector( 0,-1){ 0.5}}
%
%
%
\put( 1.5, 8.0){\oval( 1.0, 1.0)[tl]}
\put( 4.0, 8.0){\oval( 1.0, 1.0)[tr]}
\put( 1.5, 6.0){\oval( 1.0, 1.0)[bl]}
\put( 4.0, 6.0){\oval( 1.0, 1.0)[br]}
\put( 1.5, 8.5){\line( 1, 0){ 2.5}}
\put( 1.5, 5.5){\line( 1, 0){ 2.5}}
\put( 1.0, 6.0){\line( 0, 1){ 2.0}}
\put( 4.5, 6.0){\line( 0, 1){ 2.0}}
\put( 0.5, 7.5){\makebox( 4.5, 1.0)[c]{
{\color{blue}accept}
}}
\put( 0.5, 6.5){\makebox( 4.5, 1.0)[c]{
{retrieve flux tables}
}}
\put( 0.5, 6.0){\makebox( 4.5, 1.0)[c]{
{and on-the-fly products}
}}
\put( 0.5, 5.5){\makebox( 4.5, 1.0)[c]{
{from GAVO database}
}}
\put(13.50, 6.50){\vector(-1,0){ 9.0}}
%
%
%
\put( 5.5, 8.0){\oval( 1.0, 1.0)[tl]}
\put( 8.0, 8.0){\oval( 1.0, 1.0)[tr]}
\put( 5.5, 7.5){\oval( 1.0, 1.0)[bl]}
\put( 8.0, 7.5){\oval( 1.0, 1.0)[br]}
\put( 5.5, 8.5){\line( 1, 0){ 2.5}}
\put( 5.5, 7.0){\line( 1, 0){ 2.5}}
\put( 5.0, 7.5){\line( 0, 1){ 0.5}}
\put( 8.5, 7.5){\line( 0, 1){ 0.5}}
\put( 4.5, 7.5){\makebox( 4.5, 1.0)[c]{
{\color{red}request}
}}
\put( 4.5, 7.0){\makebox( 4.5, 1.0)[c]{
{\color{red} exact $T_{\mathrm{eff}}$, $\log g$, $\{X_{\mathrm{i}}\}$}
}}
\put( 8.50, 7.75){\line( 1, 0){ 1.75}}
\put(10.25, 7.75){\vector( 0, 1){ 1.75}}
%
%
%
\put(14.0,13.5){\oval( 1.0, 1.0)[tl]}
\put(16.0,13.5){\oval( 1.0, 1.0)[tr]}
\put(14.0, 5.5){\oval( 1.0, 1.0)[bl]}
\put(16.0, 5.5){\oval( 1.0, 1.0)[br]}
\put(14.0,14.0){\line( 1, 0){ 2.0}}
\put(14.0, 5.0){\line( 1, 0){ 2.0}}
\put(13.5, 5.5){\line( 0, 1){ 8.0}}
\put(16.5, 5.5){\line( 0, 1){ 8.0}}
\put(13.5,12.0){\makebox( 3.0, 3.0)[c]{
GAVO database
}}
\put(13.5,11.5){\makebox( 3.0, 3.0)[c]{
{\color{red}$T_{\mathrm{eff}}$, $\log g$, $\{X_{\mathrm{i}}\}$}
}}
\put(14.2,11.75){\oval( 1.0, 1.0)[tl]}
\put(15.8,11.75){\oval( 1.0, 1.0)[tr]}
\put(14.2,11.00){\oval( 1.0, 1.0)[bl]}
\put(15.8,11.00){\oval( 1.0, 1.0)[br]}
\put(14.2,12.25){\line( 1, 0){ 1.6}}
\put(14.2,10.50){\line( 1, 0){ 1.6}}
\put(13.7,11.00){\line( 0, 1){ 0.75}}
\put(16.3,11.00){\line( 0, 1){ 0.75}}
\put(13.5,10.5){\makebox( 3.0, 3.0)[c]{
{\color{green}ARI}
}}
\put(13.5,10.0){\makebox( 3.0, 3.0)[c]{
{\color{blue}meta data}
}}
\put(13.5, 9.5){\makebox( 3.0, 3.0)[c]{
{VO services}
}}
\put(15.0,10.50){\vector( 0,-1){ 0.75}}
\put(14.2, 9.25){\oval( 1.0, 1.0)[tl]}
\put(15.8, 9.25){\oval( 1.0, 1.0)[tr]}
\put(14.2, 6.00){\oval( 1.0, 1.0)[bl]}
\put(15.8, 6.00){\oval( 1.0, 1.0)[br]}
\put(14.2, 9.75){\line( 1, 0){ 1.6}}
\put(14.2, 5.50){\line( 1, 0){ 1.6}}
\put(13.7, 6.00){\line( 0, 1){ 3.25}}
\put(16.3, 6.00){\line( 0, 1){ 3.25}}
\put(13.5, 8.0){\makebox( 3.0, 3.0)[c]{
{\color{green}IAAT}
}}
\put(13.5, 7.5){\makebox( 3.0, 3.0)[c]{
{\color{blue}models}
}}
\put(13.5, 7.0){\makebox( 3.0, 3.0)[c]{
{atomic data}
}}
\put(13.5, 6.5){\makebox( 3.0, 3.0)[c]{
{frequency grids}
}}
\put(13.5, 6.0){\makebox( 3.0, 3.0)[c]{
{\color{blue}flux tables}
}}
\put(13.5, 5.5){\makebox( 3.0, 3.0)[c]{
\hbox{}\hspace{5.5mm}5 - 2000\,\AA}}
\put(13.5, 5.0){\makebox( 3.0, 3.0)[c]{
2000 - 3000\,\AA}}
\put(13.5, 4.5){\makebox( 3.0, 3.0)[c]{
\hbox{}\hspace{1.8mm}3000 - 55000\,\AA
}}
\end{picture}\vspace{-50mm}
  \caption{Flow diagram of TheoSSA. The VO user sends an SED request to the GAVO database by entering the
           photospheric parameters. If a suitable model is available within the desired tolerance limits, 
           it is offered as a results table. In case that the parameters are not exactly matched, the VO user
           may decide to calculate a model with the exact parameters. TMAW will start a model-atmosphere 
           calculation at our institute's (IAAT) PC cluster then. Extended model grids make use of compute resources
           that are provided by AstroGrid-D.
           As soon as the model is converged, the VO user can retrieve 
           the SED table from the GAVO database.} 
  \label{fig:theossa}
\end{figure*}
}

With the increasing usage of TheoSSA over the last years,
it became necessary to demonstrate the reliability of the SEDs.
We established simple benchmark tests \citep{ringatrauchwerner2012}
to show the achievable analysis precision, e.g\@. in the determination of
effective temperatures (\Teff) and surface gravities (\logg), in cases that TMAW SEDs
are used which are calculated with standard model atoms that are limited in
the number of atomic levels treated in NLTE. This guarantees model calculations
in a reasonable time for a VO user.

Since 2012 TheoSSA also includes synthetic spectra of spectrophotometric
standard stars. In this paper, we start to systematically establish a database
of these and address the reliability of state-of-the-art
model-atmosphere spectra and the achievable limits in future flux calibration.

\begin{table}[ht!]\centering
\caption{Parameters of the HST DA standard stars \citep{gianninasetal2011}.}
\label{tab:gianninas}
\begin{tabular}{llr@{\,$\pm$\,}lr@{\,$\pm$\,}l}
\hline
\noalign{\smallskip}
&& 
\multicolumn{2}{c}{$T_\mathrm{eff}$} & 
\multicolumn{2}{c}{$\log g$} \vspace{-2mm}\\
name & WD no.\tablefootmark{a} &
\multicolumn{2}{c}{}& 
\multicolumn{2}{c}{}\vspace{-2mm}\\
&&
\multicolumn{2}{c}{[K]} & 
\multicolumn{2}{c}{[$\mathrm{cm/sec^2}$]} \\
\noalign{\smallskip}
\hline
\noalign{\smallskip}
\gb                               & 0501+527 & 60\,920 &  \hbox{}\hspace{2mm}993                  & \hbox{}\hspace{1mm}7.55 & 0.05 \\
\object{GD\,71}                   & 0549+158 & 33\,590 &  \hbox{}\hspace{2mm}483                  &                    7.93 & 0.05 \\
\object{GD\,153}                  & 1254+223 & 40\,320 &  \hbox{}\hspace{2mm}626                  &                    7.93 & 0.05 \\ 
\object{HZ\,43A}\tablefootmark{b} & 1314+293 & 56\,800 &                    1249\tablefootmark{c} &                    7.89 & 0.07 \\
\hline
\end{tabular}
\tablefoot{~\\
\tablefoottext{a}{WD numbers are from \citet{mccooksion1999}.}
\tablefoottext{b}{\object{HZ\,43A} is only used in the UV because of contamination at longer wavelengths from its M-dwarf companion.}
\tablefoottext{c}{\citet{beuermannetal2006, beuermannetal2008}
                  determined \Teffw{51\,111 \pm 660} and \loggw{7.90 \pm 0.080}.}
}
\end{table}

White dwarfs (WDs) are ideal objects for the calibration of  astronomical observations \citep{rauch2013}. 
They are relatively simple objects and their radiation is determined by fundamental physics, 
e.g\@. their radius is defined by electron degeneracy. Moreover, they are nearby and their distance
can be measured precisely, at least by the upcoming 
GAIA\footnote{\url{http://www.esa.int/Our_Activities/Space_Science/Gaia_overview}} mission
\citep[cf\@.][for a description of the GAIA spectrophotometric standard stars survey]{pancinoetal2012}.
Most of the hot, hydrogen-rich WDs (spectral type DA) with \Teff < 40000\,K have virtually pure hydrogen 
atmospheres (gravitational settling), while the hotter WDs exhibit lines of heavier elements due to
radiative levitation. WD spectral modeling requires adequate observations (WDs are intrinsically faint)
and state-of-the-art theoretical model atmospheres that account for reliable physics and deviations from local 
thermodynamic equilibrium (LTE).

The hot DA-type WD \gb (\object {BD+52$\degr$913}) is, together with
\object{GD\,71}, and 
\object{GD\,153}, 
one of the primary flux reference standards for all absolute calibrations
from 1000 to 25\,000\,\AA\ \citep{bohlin2007}. Recent results for their \Teff and \logg 
are summarized in Table\,\ref{tab:gianninas}.
\gb, the hottest and visually brightest \citep[$m_\mathrm{V} = 11.7228$, ][]{vanleeuwen2007} isolated WD 
\citep[with a well known distance of 57.96\,pc, ][]{andersonfrancis2012}
of the sample, is ideal for panchromatic calibration 
from the ultraviolet (UV) to the infrared (IR) wavelength range. 
However, due to its high \Teff and relatively low
\logg, radiative levitation competes against gravitational settling and holds trace elements in the 
photosphere and exhibits many weak metal lines \citep[e.g\@.][]{barstowetal2003}
in its observed UV spectrum.

\begin{table*}[ht!]\centering
\caption{\Teff and \logg from previous analyses of \gb.}
\label{tab:previous}
\setlength{\tabcolsep}{.5em}
\begin{tabular}{r@{\,$\pm$\,}lr@{\,$\pm$\,}lll}
\hline
\noalign{\smallskip}
\multicolumn{2}{c}{$T_\mathrm{eff}$} & 
\multicolumn{2}{c}{$\log g$} \vspace{-2mm}\\
\multicolumn{2}{c}{}& 
\multicolumn{2}{c}{}&
\multicolumn{1}{c}{reference} & \multicolumn{1}{c}{method}\vspace{-2mm}\\
\multicolumn{2}{c}{[K]} & 
\multicolumn{2}{c}{[$\mathrm{cm/sec^2}$]} \\
\noalign{\smallskip}
\hline
\noalign{\smallskip}
\multicolumn{2}{l}{61\,900}                              & \multicolumn{2}{l}{\smspl 7.5\tablefootmark{a}}  & \citet{shipman1979}    & \smspr LTE, pure H, optical colors                                   \\
56\,788 & \hbox{}\hspace{1.6mm}3336                      & 5.95  & 0.04                               & \citet{koesteretal1979}      & \smspr LTE, pure H, optical colors, \tablefootmark{b}                \\
62\,250 & \hbox{}\hspace{1.6mm}3520                      & 7.55  & 0.35                               & \citet{holbergetal1986}      & \smspr LTE, pure H, \ion{H}{i} L\,$\alpha$ line                      \\
59\,250 & \hbox{}\hspace{1.6mm}2000                      & 7.50  & 0.10                               & Kidder 1990                  & \smspr LTE, pure H, \ion{H}{i} H\,$\gamma$ and                       
                                                                                                                                                                      H\,$\delta$ lines                     \\
\multicolumn{4}{l}{}                                                                                  & cited by \citet{holbergetal1991} &                                                                  \\
\multicolumn{2}{l}{61\,000 $^{+6000}_{-4000}$}           & \multicolumn{2}{l}{\smspl 8.00\tablefootmark{a}} & \citet{greenetal1990}  & \smspr LTE, H+He, extreme ultraviolet (EUV) continuum                \\
\noalign{\smallskip}                                                                                                                                                                                        
\multicolumn{2}{l}{61\,170 $^{+4830}_{-4230}$}           & \multicolumn{2}{l}{\smspl 8.00\tablefootmark{a}} & \citet{finleyetal1990} & \smspr LTE, H+He, ultraviolet (UV) continuum                         \\
62\,250 & \hbox{}\hspace{1.6mm}1000                      & \multicolumn{2}{l}{\smspl 8.00\tablefootmark{a}} & \citet{finleyetal1990} & \smspr LTE, H+He,     \ion{H}{i} L\,$\alpha$ line                    \\
\multicolumn{2}{l}{62\,250}                              & \multicolumn{2}{l}{\smspl 7.55}            & \citet{vennesetal1991}       & NLTE, H+HeCNSi, UV spectrum                                          \\
53\,500 & \hbox{}\hspace{3.9mm}500                       & \multicolumn{2}{l}{}                       & \citet{koesterfinley1992}    & \smspr LTE, H+He\tablefootmark{c}, UV continuum                      \\
60\,500 & \hbox{}\hspace{3.9mm}900                       & 7.50  & 0.05                               & \citet{vidalmadjaretal1994}  & \smspr LTE, H+He, \ion{H}{i} Balmer lines                            \\
57\,900 & \hbox{}\hspace{3.9mm}1500                      & \multicolumn{2}{l}{\smspl 7.50\tablefootmark{a}} & \citet{dupuisetal1995} & NLTE, pure H, EUV continuum\tablefootmark{d}                         \\
54\,000 & \hbox{}\hspace{3.9mm}800                       & \multicolumn{2}{l}{\smspl 7.50\tablefootmark{a}} & \citet{dupuisetal1995} & NLTE, H+CNOFe, EUV continuum\tablefootmark{d}                        \\
60\,500 & \hbox{}\hspace{1.6mm}1000                      & \multicolumn{2}{l}{\smspl 7.5}             & \citet{lanzetal1996}         & \smspr LTE, H+HeC, \ion{H}{i} Balmer lines                           \\
56\,000 & \hbox{}\hspace{1.6mm}1000                      & \multicolumn{2}{l}{\smspl 7.5}             & \citet{lanzetal1996}         & NLTE, H+HeC, \ion{H}{i} Balmer lines                                 \\
55\,200 & \hbox{}\hspace{1.6mm}1000                      & \multicolumn{2}{l}{\smspl 7.5}             & \citet{lanzetal1996}         & NLTE, H+HeCFe, \ion{H}{i} Balmer lines                               \\
64\,000 & \hbox{}\hspace{1.6mm}1000                      & 7.64  & 0.06                               & \citet{vennesetal1996}       & \smspr LTE, pure H, \ion{H}{i} Lyman lines                           \\
57\,900 & \hbox{}\hspace{1.6mm}1500                      & \multicolumn{2}{l}{\smspl 7.5}             & \citet{vennesetal1996}       & \smspr LTE, pure H, extreme UV (EUV) continuum\tablefootmark{d}      \\
64\,100 & \hbox{}\hspace{3.9mm}700                       & 7.69  & 0.04                               & \citet{vennesetal1996}       & \smspr LTE, H+CNOFe, \ion{H}{i} Balmer lines                         \\
52\,600 & \hbox{}\hspace{3.9mm}800                       & 7.53  & 0.07                               & \citet{vennesetal1996}       & \smspr LTE, H+CNOFe, \ion{H}{i} Lyman lines                          \\
54\,000 & \hbox{}\hspace{3.9mm}800                       & \multicolumn{2}{l}{\smspl 7.5}             & \citet{vennesetal1996}       & \smspr LTE, H+CNOFe, EUV continuum\tablefootmark{d}                  \\
61\,193 & \hbox{}\hspace{3.9mm}241                       & 7.49  & 0.01                               & \citet{finleyetal1997}       & \smspr LTE, H-Ni, \ion{H}{i} Balmer lines                            \\
\multicolumn{2}{l}{59\,160 $^{+1270}_{-1070}$}           & \multicolumn{2}{l}{7.36$^{+0.08}_{-0.07}$} & \citet{barstowetal1998}      & NLTE, pure H, \ion{H}{i} Balmer lines                                \\
\noalign{\smallskip}                                                                                                                                                                                        
\multicolumn{2}{l}{59\,190 $^{+1400}_{-\,\,\,820}$}      & \multicolumn{2}{l}{7.36$^{+0.07}_{-0.07}$} & \citet{barstowetal1998}      & NLTE, H+He, \ion{H}{i} Balmer lines                                  \\
\noalign{\smallskip}                                                                                                                                                                                        
\multicolumn{2}{l}{59\,060 $^{+1130}_{-1090}$}           & \multicolumn{2}{l}{7.36$^{+0.08}_{-0.07}$} & \citet{barstowetal1998}      & NLTE, H+He + heavy-metal poor, \ion{H}{i} Balmer lines               \\
\noalign{\smallskip}                                                                                                                                                                                        
\multicolumn{2}{l}{53\,840 $^{+\,\,\,400}_{-\,\,\,160}$} & \multicolumn{2}{l}{7.38$^{+0.07}_{-0.08}$} & \citet{barstowetal1998}      & NLTE, H+He + heavy-metal rich, \ion{H}{i} Balmer lines               \\
\noalign{\smallskip}                                                                                                                                                                                        
52\,920 & \hbox{}\hspace{3.9mm}350                       & 7.36  & 0.03                               & \citet{barstowetal1998}      & NLTE, H+He + heavy-metal rich, \ion{H}{i}                            
                                                                                                                                                                      Lyman lines\tablefootmark{e}          \\
\multicolumn{2}{l}{56\,000}                              & \multicolumn{2}{l}{\smspl 7.6}             & \citet{wolffetal1998}        & \smspr LTE + NLTE, H+CNOSiFeNi, EUV continuum\tablefootmark{d}       \\
\multicolumn{2}{l}{56\,000}                              & \multicolumn{2}{l}{\smspl 7.6}             & \citet{dreizlerwolff1999}    & NLTE, H+CNOSiFeNi, diffusion model, EUV to optical                   \\
54\,600 & \hbox{}\hspace{3.9mm}200                       & 7.60  & 0.02                               & \citet{barstowetal2001}      & NLTE, H+HeCNOSiFeNi, \ion{H}{i} Balmer lines                         \\
52\,930 & \hbox{}\hspace{1.6mm}3600                      & 7.16  & 0.2                                & \citet{barstowetal2001}      & NLTE, H+HeCNOSiFeNi, \ion{H}{i} Lyman lines\tablefootmark{f}         \\
53\,180 & \hbox{}\hspace{3.9mm}530                       & 7.43  & 0.04                               & \citet{barstowetal2001}      & NLTE, H+HeCNOSiFeNi, \ion{H}{i} Lyman lines\tablefootmark{e}         \\
\multicolumn{2}{l}{56\,000}                              & \multicolumn{2}{l}{\smspl 7.59}            & \citet{schuhetal2002}        & NLTE, H+HeCNOSiFeNi, diffusion model, EUV continuum\tablefootmark{d} \\
\multicolumn{2}{l}{54\,000}                              & \multicolumn{2}{l}{\smspl 7.5}             & \citet{holbergetal2003}      & NLTE, metal lines                                                    \\
58\,865 & \hbox{}\hspace{3.9mm}706                       & 7.57  & 0.038                              & \citet{lajoiebergeron2007}   & NLTE\tablefootmark{g}, pure H, \ion{H}{i} Balmer lines               \\
60\,680 &                     15\,000   & \multicolumn{2}{l}{\smspl 7.57\tablefootmark{h}} & \citet{lajoiebergeron2007}   & NLTE\tablefootmark{g}, pure H, \ion{H}{i} Lyman lines\tablefootmark{i}          \\
57\,414 & \hbox{}\hspace{1.6mm}4700     & \multicolumn{2}{l}{\smspl 7.57\tablefootmark{h}} & \citet{lajoiebergeron2007}   & NLTE\tablefootmark{g}, pure H, V-normalization method                           \\
61\,980 & \hbox{}\hspace{4.0mm}514                       & 7.56  & 0.04                               & \citet{allendeprietoetal2009}& NLTE, H\tablefootmark{j}, \ion{H}{i} Balmer lines                    \\
60\,920 & \hbox{}\hspace{4.0mm}993                       & 7.55  & 0.05                               & \citet{gianninasetal2011}    & NLTE, H+CNO, \ion{H}{k} Balmer lines\tablefootmark{k}                \\
\hline
\end{tabular}
\tablefoot{~\\
\tablefoottext{a}{Assumed \logg value.}
\tablefoottext{b}{The authors note that the results are extrapolated from their model grid.}
\tablefoottext{c}{Stratified model, hydrogen-layer mass between $6\times10^{-15}$ and $8\times10^{-15}$\,\Msol.}
\tablefoottext{d}{Extreme Ultraviolet Explorer (EUVE, \url{http://heasarc.gsfc.nasa.gov/docs/euve/euve.html}) observations.}
\tablefoottext{e}{Orbiting and Retrievable Far and Extreme Ultraviolet Spectrometer (ORFEUS, \url{http://www.uni-tuebingen.de/en/4221}) and FUSE observations.}
\tablefoottext{f}{Hopkins Ultraviolet Telescope (HUT, \url{http://praxis.pha.jhu.edu/}) observations.}
\tablefoottext{g}{International Ultraviolet Explorer (IUE) observations.}
\tablefoottext{h}{Adopted from their optical solution.}
\tablefoottext{i}{Models described in \citet{liebertetal2005}.}
\tablefoottext{j}{The authors note that \Teff may be overestimated by $\approx 6\,000$\,K because their pure-H models are inappropriate due to the photospheric metal content.}
\tablefoottext{k}{New \ion{H}{i} Stark line-broadening tables from \citet{tremblaybergeron2009}.}
}
\end{table*}

A variety of previous spectral analyses of \gb (Table\,\ref{tab:previous})
had shown that it is difficult to determine its \Teff precisely. 
\citet{barstowetal1998} found that the metal content in the photosphere has a strong
impact on the determined \Teff. \Teffw{60\,920} was found by the most recent analysis \citep{gianninasetal2011}
who considered only C, N, and O (at solar abundances) in their models. The neglect of other metals
calls for improved models with better metal opacities. The same may be true for HZ\,43A even if metals are below
the detection limit of the available spectra (Table\,\ref{tab:gianninas}).

Many abundance analyses were performed, most of them \citep[e.g\@.][]{barstowetal2005}, were based on 
previous \Teff determinations from Balmer-line fits (cf\@. Table\,\ref{tab:previous}) and not from
self-consistent fits to models with varying metal abundances.
\citet{lanzetal1996} measured He, C, N, O, Si, Fe, and Ni abundances,
\citet{holbergetal2003} determined abundances of C, N, O, Al, Si, Fe, and Ni
and gave upper limits for Mg, Cr, Mn, and Co.
The compiled abundances are listed in in Table\,\ref{tab:prevabund}.

\begin{table}[ht!]\centering
\caption{Abundances of photospheric trace elements in \gb from previous analyses.
         Table\,\ref{tab:previous} displays the \Teff and \logg values of the employed models.}
\label{tab:prevabund}
\begin{tabular}{ccr@{.}lr@{.}l}
\hline
\hline
\noalign{\smallskip}
element && \multicolumn{4}{c}{log mass fraction} \\ 
\cline{1-1}
\cline{3-6}
\noalign{\smallskip}
He && $  -4$&$2\pm 0.1\tablefootmark{a} $ & $  -4$&$4\tablefootmark{i} $ \\
   && $  -4$&$4\pm 0.3\tablefootmark{b} $ & \multicolumn{2}{c}{}         \\
C  && $  -4$&$6\pm 0.3\tablefootmark{c} $ & $  -5$&$6\tablefootmark{i} $ \\
   && $  -4$&$6\pm 0.3\tablefootmark{b} $ & \multicolumn{2}{c}{}         \\
N  && $  -4$&$3\pm 0.4\tablefootmark{d} $ & $  -5$&$9\tablefootmark{i} $ \\
   && $  -5$&$6\pm 0.3\tablefootmark{b} $ & \multicolumn{2}{c}{}         \\
O  && $  -4$&$8\pm 0.3\tablefootmark{e} $ & $  -4$&$6\tablefootmark{i} $ \\
   && $  -4$&$8\pm 0.3\tablefootmark{b} $ & \multicolumn{2}{c}{}         \\
Mg && \multicolumn{2}{c}{}                & $ <-5$&$6\tablefootmark{i} $ \\
Al && \multicolumn{2}{c}{}                & $  -5$&$1\tablefootmark{i} $ \\
Si && $  -5$&$1\pm 0.4\tablefootmark{d} $ & $  -5$&$0\tablefootmark{i} $ \\
   && $  -5$&$1\pm 0.5\tablefootmark{c} $ & \multicolumn{2}{c}{}         \\
   && $  -5$&$0\pm 0.3\tablefootmark{b} $ & \multicolumn{2}{c}{}         \\
P  && $  -6$&$2\pm 0.2\tablefootmark{c} $ & \multicolumn{2}{c}{}         \\
S  && $  -5$&$2\pm 0.5\tablefootmark{c} $ & \multicolumn{2}{c}{}         \\
Cl && $< -7$&$0$\tablefootmark{b}         & \multicolumn{2}{c}{}         \\
Cr && \multicolumn{2}{c}{}                & $ <-6$&$3\tablefootmark{i} $ \\
Mn && \multicolumn{2}{c}{}                & $ <-6$&$3\tablefootmark{i} $ \\
Fe && $  -3$&$8\pm 0.3\tablefootmark{d} $ & $  -3$&$8\tablefootmark{i} $ \\
   && $  -3$&$4\pm 0.4\tablefootmark{f} $ & \multicolumn{2}{c}{}         \\
   && $  -3$&$3\pm 0.3\tablefootmark{b} $ & \multicolumn{2}{c}{}         \\
Co && \multicolumn{2}{c}{}                & $ <-6$&$2\tablefootmark{i} $ \\
Ni && $  -4$&$2\pm 0.5\tablefootmark{g} $ & $  -4$&$4\tablefootmark{i} $ \\
   && $  -4$&$2\pm 0.4\tablefootmark{f} $ & \multicolumn{2}{c}{}         \\
   && $  -3$&$9\pm 0.3\tablefootmark{b} $ & \multicolumn{2}{c}{}         \\
Ge && $  -6$&$1\pm 0.2\tablefootmark{h} $ & \multicolumn{2}{c}{}         \\
Sn && $  -6$&$9\pm 0.2\tablefootmark{h} $ & \multicolumn{2}{c}{}         \\
\hline
\end{tabular}
\tablefoot{~\\
\tablefoottext{a}{\citet{cruddaceetal2002}}
\tablefoottext{b}{\citet{lanzetal1996}}
\tablefoottext{c}{\citet{vennesetal1996}}
\tablefoottext{d}{\citet{vidalmadjaretal1994}}
\tablefoottext{e}{\citet{chayeretal1996}}
\tablefoottext{f}{\citet{wernerdreizler1994}}
\tablefoottext{g}{\citet{holbergetal1994}}
\tablefoottext{h}{\citet{vennesetal2005}}
\tablefoottext{i}{\citet{holbergetal2003}, He abundance assumed, no abundance uncertainties given}
}
\end{table}

Based on a grid of state-of-the-art line-blanketed NLTE model atmospheres that include opacities of all identified
metals, we perform a detailed spectral analysis. We describe the available observations in Sect.\,\ref{sect:obs},
followed by a brief introduction to our model atmosphere code and the atomic data (Sect.\,\ref{sect:models}).
The spectral analysis is summarized in Sect.\,\ref{sect:analysis} and we end with our conclusions
(Sect.\,\ref{sect:conclusions}).

\section{Observations}
\label{sect:obs}

\subsection{FUSE data}
\label{sect:fusedata}

\gb was observed many times over the course of the FUSE mission in the wavelength range $910\,\mathrm{\AA} -  1190\,\mathrm{\AA}$,
both for calibration purposes and for studies of the interstellar medium.
For the present study, only observations obtained in the first eight months of the mission through the LWRS spectrograph aperture were analyzed. 
This time period included the majority of the LWRS exposure time obtained during the mission, and had the secondary benefit that the detectors 
had not yet suffered much degradation from gain sag.
The observation IDs of the datasets were: M1010201, M1010202, M1030602, P1041203, and S3070101.

Apart from a few quirks affecting the M1010201 and M1030602 observations, which were among the first obtained during the mission, the quality 
of the data is excellent. No SiC data were obtained in observations M1010201 or M1030602 as a result of channel mis-alignment. Otherwise: 
exposure-to-exposure variations in flux were typically well under 1\,\%, indicating good channel alignment. 
The detector region used to record spectral image data for LiF2b was offset from the actual spectrum position
in the M1010201 observation, so those spectra were discarded. 
The net exposure times were 33.3\,ksec for the 
SiC channels, 36\,ksec for LiF2b, and 40\,ksec for LiF1 and LiF2a.

No special processing was applied to data from individual exposures. Raw data were processed with CalFUSE v3.2.3.
Zero-point offsets in the wavelength scale were adjusted for each exposure by shifting each spectrum   to coalign narrow interstellar absorption features. 
In order to assess the influence of geocoronal airglow emission, spectra obtained during orbital day and night were combined separately.
All the observations were obtained in spectral image (``histogram'') mode, so no information on photon arrival time was available within an exposure. 
However, the timeline table in the intermediate data files was examined for each exposure to determine the time spent in the ``Day'' and ``Night'' portions of the orbit. 
If the ``Day'' portion of such an exposure exceeded 15\,\% of the total exposure duration, it was included with the other Day spectra.  
Because histogram mode exposures were short, most exposures were entirely Day or entirely Night.

The individual exposures from all five observations were then combined to form composite Day and Night spectra for each channel. 
The Day and Night spectra were then compared at the locations of all the known airglow emission lines. 
If the Day spectra showed any excess flux in comparison to the Night spectra at those locations, the corresponding pixels in the Day spectra were 
flagged as bad and were not included in subsequent processing. Significant airglow emission during orbital Day was seen for most observations at 
\ion{H}{i} Ly\,$\beta$ through Ly\,$\delta$, and \Ionww{O}{1}{988, 1027, 1028, 1039}. 
Significant airglow was present during orbital night only at Ly\,$\beta$; 
this affects the interstellar-absorption profile but has no effect on our analysis of the photospheric spectrum.

The final step was to combine the spectra from the four instrument channels into a single composite spectrum. Because of residual distortions in the 
wavelength scale in each channel, additional shifts of localized regions of each spectrum were required to coalign the spectra; such shifts were typically 
only one or two pixels. Bad pixels resulting from detector defects were flagged at this point and excluded from further processing. 
Finally, the spectra were resampled onto a common wavelength scale and combined, weighting by signal to noise on a pixel-by-pixel basis.

The signal to noise of the final combined spectrum is limited by fixed-pattern noise in the detectors. The final spectrum has a minimum of roughly 
20\,000 counts per 0.013\,\AA\ pixel in the continuum, near the Lyman edge, and 60\,000 - 130\,000 counts per pixel long-ward of 1000\,\AA. 
The effects of fixed-pattern noise are minimized by the fact that the positions of the spectra on the detectors varied during each observation, 
and by the fact that nearly every wavelength bin was sampled by at least two different detectors.

\subsection{HST data}
\label{sect:hstdata}

As described in detail by Bohlin \& Gordon (in prep.), the HST/STIS low-dispersion
flux calibration is derived from an ensemble match to the NLTE TLUSTY (version
203) model atmosphere SEDs for pure hydrogen \citep{hubenylanz1995}. The models are
for \gb, \object{GD\,71}, and \object{GD\,153}. Originally, \object{HZ\,43A} was also used as a standard star
but fell off the list of primary flux standards because of an M star companion
that contaminates the STIS observations in the visible and IR \citep{bohlinetal2001}.

For the STIS \'echelle modes, the flux calibration is based only on the TLUSTY
model for \gb. The \'echelle absolute fluxes are less precise than for low
dispersion because of the single model for the reference fluxes, because of
imprecision in the matching of the separate echelle orders, and because the
plethora of weak lines at the shorter wavelengths are missing in the reference
SED. However, the STIS echelle narrow metal line profiles are unaffected by
uncertainties in the absolute fluxes.

For the highest STIS resolution of $\approx 3$\,km/s, there are two modes, namely E140H and E230H,
which require several central wavelength settings for complete wavelength
coverage from $1145\,-\,3145$\,\AA. Because \gb is the primary STIS \'echelle
calibration star with repeated observations, 105 observations in the $0\farcs 2\times 0\farcs 2$
aperture are available from the Mikulski Archive for Space Telescopes
(MAST)\footnote{\url{http://archive.stsci.edu/}}. Each spectrum is resampled to a
wavelength grid with a sampling interval corresponding to a resolving power of $R=2.3 \times 10^5$ and co-added.
The number of individual observations at each wavelength point ranges from $4 - 44$,
while the total exposure time ranges from $6400 - 64\,000$\,s at each point. The total
counts in electrons at each point in the continuum are typically well above 1000
and range up to over 10\,000 from $1225 -1400$\,\AA, where the statistical uncertainty
is sometimes better than 1\,\%. The high-dispersion \'echelle spectrum is available
from the CALSPEC\footnote{\url{http://www.stsci.edu/hst/observatory/cdbs/calspec.html}}
database along with the STIS low-dispersion data.

The photospheric radial velocity $v_\mathrm{rad} = 22.1\pm 0.6\,\mathrm{km/s}$ measured by \citet{holbergetal1994} 
matches well our STIS observation. We adopt this value for our analysis.

\subsection{Interstellar line absorption and reddening}
\label{sect:ism}

\begin{figure}
  \resizebox{\hsize}{!}{\includegraphics{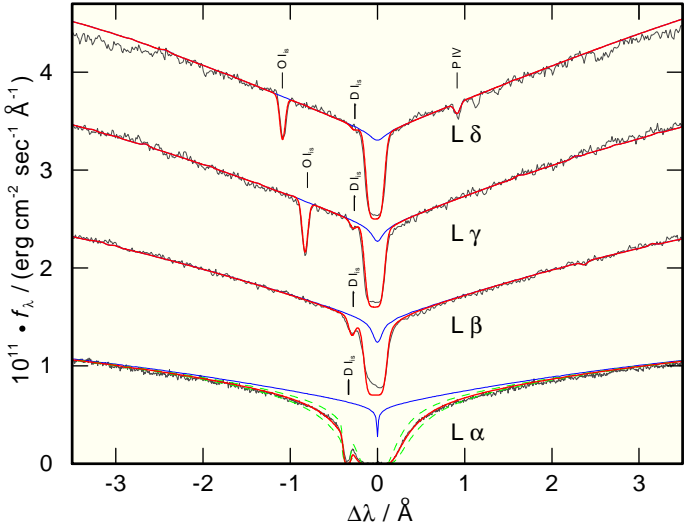}}
  \caption{Comparison of STIS and FUSE observations around \ion{H}{i} L\,$\alpha$ - $\delta$ 
           with our final model. 
           Thick (red in the online version) photospheric + ISM line absorption model with $N_\ion{H}{i} = 2.2\times 10^{18}\,\mathrm{cm^{-2}}$,
           thin (blue) pure photospheric model;
           dashed (green, L\,$\alpha$ only) $N_\ion{H}{i} = 1.2\times 10^{18}\,\mathrm{cm^{-2}}$, $3.2\times 10^{18}\,\mathrm{cm^{-2}}$. 
           The locations of the \ion{D}{i} blends are marked. 
           L\,$\beta$ - $\delta$ are shifted in flux (0.7, 1.6, $2.7\times 10^{-11}\,\mathrm{(erg\,cm^{-2}\,s^{-1}\,\AA^{-1})}$ for clarity.
           A reddening of $E_{B-V} = 0.0005$ is applied following the law of \citet[][with $R_V = 3.1$]{fitzpatrick1999}.
          } 
  \label{fig:nhi}
\end{figure}

The interstellar neutral hydrogen density $N_\ion{H}{i}$ was determined from the comparison
of our final model with the STIS and FUSE observations (Fig.\,\ref{fig:nhi}).
In all plots shown in this paper, we modeled the interstellar medium (ISM) line absorption 
(using Voigt line profiles) with
WRPLOT\footnote{\url{http://www.astro.physik.uni-potsdam.de/~htodt/wrplot/index.html}}.
The best match is found for $\log\ (N_\ion{H}{i}\,/\,\mathrm{cm^2}) = 18.34^{+0.08}_{-0.10}$.
The \ion{D}{i} blends to \ion{H}{i} L\,$\alpha$ - $\delta$ are clearly visible and
best reproduced at $\log\ (N_\ion{D}{i}\,/\,\mathrm{cm^2}) = 13.54^{+0.05}_{-0.06}$, i.e\@.
D/H = $1.59^{+0.41}_{-0.65}\times 10^{-5}$,
Our values are in good agreement with those determined by \citet{lemoineetal2002},
$\log\ (N_\ion{H}{i}\,/\,\mathrm{cm^2}) = 18.18\pm 0.18$ and
D/H = $1.66^{+0.9}_{-0.6}\times 10^{-5}$ (both with 2\,$\sigma$ errors).

\begin{figure}
  \resizebox{\hsize}{!}{\includegraphics{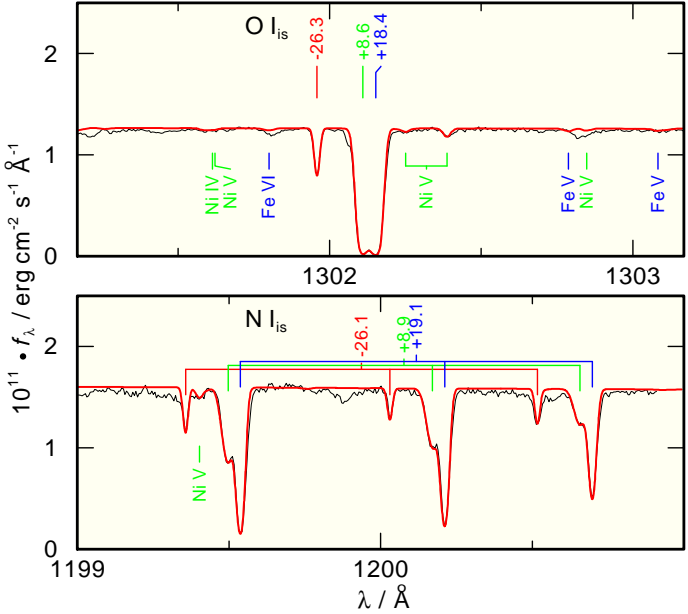}}
  \caption{STIS spectrum around the interstellar absorption lines
          \ion{O}{i} $\lambda\, 1302.163\,\mathrm{\AA}$ (top) and
          \ion{N}{i} $\lambda\lambda\, 1199.550, 1200.223, 1200.710\,\mathrm{\AA}$ compared with
          the synthetic spectrum of our final model where the ISM lines are included.
          The labels give the radial velocities (in km/sec) that are applied. 
          } 
  \label{fig:nioi}
\end{figure}

Besides \ion{H}{i} and \ion{D}{i}, we identified interstellar lines of
\ion{C}{ii}  - \ion{}{iv},
\ion{N}{i}   - \ion{}{ii},
\ion{O}{i},
\ion{Al}{ii},
\ion{Si}{ii} - \ion{}{iii},
\ion{P}{i}   - \ion{}{ii},
\ion{S}{i}   - \ion{}{ii}, and
\ion{Fe}{ii}
in the FUSE and STIS spectra (Table\,\ref{tab:lineids}).
To identify pure photospheric lines that are contaminated by ISM lines,
we modeled all of these and found that we need two distinct
clouds with $v_\mathrm{rad} = 9 \pm 1\,\mathrm{km/s}$ and $v_\mathrm{rad} = 19 \pm 1\,\mathrm{km/s}$.
This is well in agreement with the measurements of 
\citet[$8.6 \pm 1.7\,\mathrm{km/s}$ and $19.3 \pm 2.5\,\mathrm{km/s}$][]{sahuetal1999}, 
who assigned the latter value to the local interstellar cloud.
\citet{dickinsonetal2012} measured $8.5 \pm 0.18\,\mathrm{km/s}$ and $19.3 \pm 0.03\,\mathrm{km/s}$.
They unambiguously detected that the first cloud is of circumstellar origin.
An additional third cloud with intermediate velocity like assumed by 
\citet[$v_\mathrm{rad} = 8.2, 13.2, 20.3 \,(\pm 0.8\,\mathrm{km/s}$]{vidalmadjaretal1998}
is not necessary for our modeling (Fig.\,\ref{fig:nioi}, top).

Interestingly, we find additional weak absorptions of \Ionw{O}{1}{1302.163} and
\Ionw{N}{1}{1199.550, 1200.223, 1200.710} at $v_\mathrm{rad}$ of $-26.3,\mathrm{km/s}$ and
$-26.1,\mathrm{km/s}$, respectively. These velocities are reminiscent of the expansion velocity of a planetary
nebula shell \citep[e.g.\@.][]{kwoketal1978}, that for a stellar mass of  $M = 0.555\,M_\odot$ (Sect.\,\ref{sect:distance})
must have been ejected more than 500\,000 years ago \citep{renedoetal2010}. Its recombined, neutral gas, however, is still 
in the line of sight.

\begin{figure}
  \resizebox{\hsize}{!}{\includegraphics{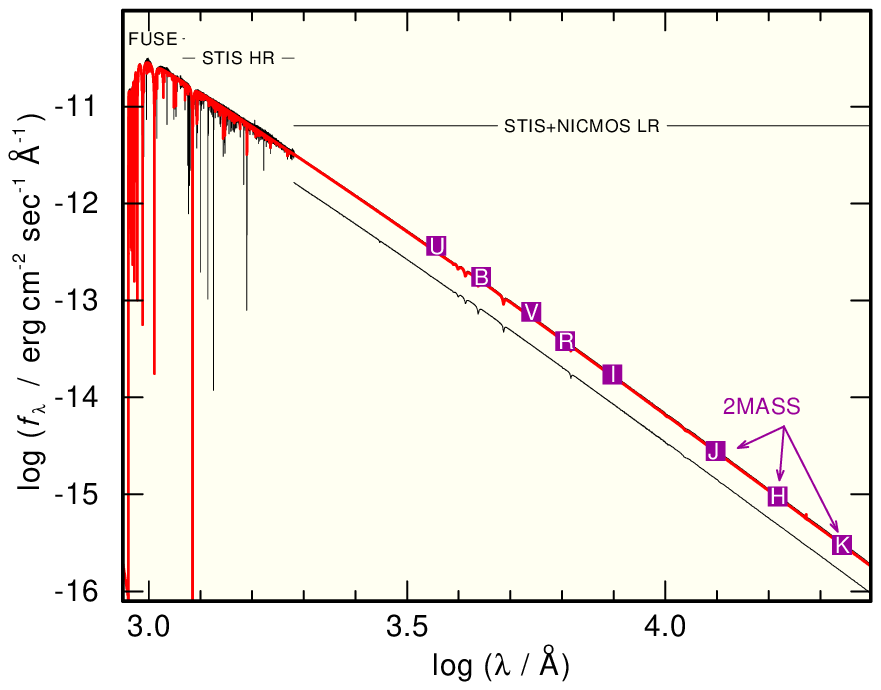}}
  \caption{Comparison of FUSE and HST (STIS and NICMOS) observations with our final model.
           A reddening of $E_{B-V} = 0.0005$ is applied. The low-resolution (LR) STIS+NICMOS
           observation vanishes behind the model SED due to the line width.
           Therefore, we plotted the observed spectrum twice, one shifted by $\Delta \log f_\lambda = -0.2$
           for clarity.
           U, R, I \citep{landoltuomoto2007},
           B, V \citep{hogetal2000}, J, H, and K \citep{cutrietal2003} fluxes 
           \citep[converted from brightnesses using values given by][]{heberetal2002} are
           shown for comparison.
          } 
  \label{fig:ebv}
\end{figure}

From the low interstellar $N_\ion{H}{i}$ density, we expect a low interstellar reddening.
The Galactic reddening law of \citet[][]{groenewegenlamers1989}, $\log (N_\ion{H}{i}/E_{B-V}) = 21.58\pm 0.10$, predicts
$0.0003\,\sla\,E_{B-V}\,\sla\,0.0007$. Figure\,\ref{fig:ebv} shows a comparison of observations
and synthetic spectrum from the far UV (FUV) to the IR. We find $E_{B-V} = 0.0005 \pm 0.0005$.

\section{Model atmospheres and atomic data}
\label{sect:models}

Table\,\ref{tab:previous} demonstrates clearly that a panchromatic analysis from
the EUV to the optical is inevitable for accurate results on photospheric parameters. 
Moreover, NLTE modeling is mandatory to calculate a reliable synthetic spectrum. 

\citet{lanzetal1996} presented the first NLTE model (Table\,\ref{tab:previous}) that
reproduced the observed spectrum from the EUV to the optical wavelength range.
\citet{barstowhubeny1998} introduced then a stratified H+He envelope including heavier metals
in their models to improve the match to the observed flux below the \ion{He}{ii} absorption
threshold ($\lambda\, \la\, 228$\,\AA). Later analyses had shown that there is further evidence for a 
stratification in \gb's photosphere. \citet{vennesetal2000} closely examined \object{Feige\,24} that, 
compared with \gb, has \emph{similar atmospheric parameters and an almost identical abundance
pattern}. They found that the \ion{O}{iv}\,/\,\ion{O}{v} ionization equilibrium is overcorrected 
by $-0.8$\,dex in their NLTE model. They concluded that this \emph{might reveal an inhomogeneous 
vertical stratification of oxygen} in both stars. A later analysis of both stars \citep{venneslanz2001} 
showed that the average heavy-metal abundance in \object{Feige\,24} is 0.17\,dex larger compared to 
the cooler and, hence, older \gb (same \logg). Thus, the abundance pattern is determined by the same 
processes in both stars and the authors assumed that selective radiative pressure and gravity are in 
diffusive equilibrium. This was proven by
\citet{dreizlerwolff1999}. They used self-consistent diffusion models (Table\,\ref{tab:previous})
that were able to reproduce the observed flux for $\lambda\, \la\, 228$\,\AA\ without additional
absorbers or mechanisms. However, problems remained with the fit to the UV lines.

Now, our strategy to proceed with the analysis is threefold. We start with chemically homogeneous 
models to find the model that reproduces best the continuum slope and the spectral lines from 
the FUV to the optical (Sect.\,\ref{sect:analysis}).
In an intermediate step, we will then apply the depth-dependent abundance profiles calculated by 
\citet{dreizlerwolff1999} to our final homogeneous model to investigate the impact of chemical 
stratification on the emergent spectrum (Sect.\,\ref{sect:profile}). In the last step, a diffusion 
model is calculated and compared with the homogeneous model (Sect.\,\ref{sect:diffusion}).

\section{Spectral analysis and results}
\label{sect:analysis}

The metal-line blanketed NLTE model atmospheres for our analysis were calculated with the
state-of-the-art T\"ubingen NLTE model-atmosphere package 
\citep[][TMAP\footnote{\url{http://astro.uni-tuebingen.de/~TMAP}}]{werneretal2003},
which can consider opacities of all elements from H to Ni and beyond 
\citep{rauch1997,rauch2003,werneretal2012,rauchetal2012}.
TMAP was successfully used for the spectral analysis of hot, compact stars
\citep[e.g\@.][]{rauchetal2007,wassermannetal2010,ziegleretal2012}.

\begin{figure}
  \resizebox{\hsize}{!}{\includegraphics{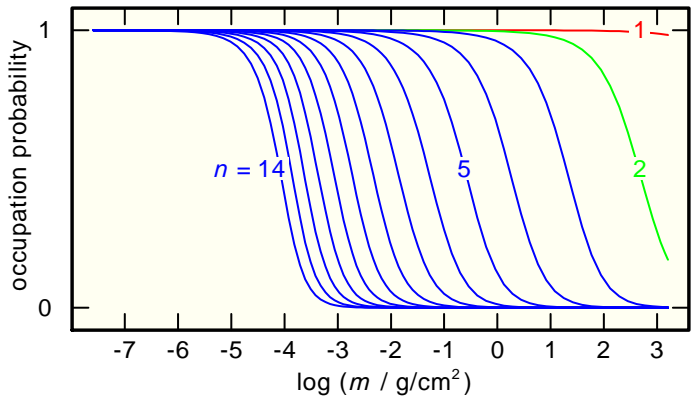}}
  \caption{Occupation probabilities of the \ion{H}{i} levels
           with principal quantum numbers $n$ = 1 $-$ 14 in our
           final model.            
          } 
  \label{fig:HIoccu}
\end{figure}

Our models assume plane-parallel geometry and are in hydrostatic and radiative
equilibrium. Opacities of all species for which spectral lines are identified, namely 
H,
He,
C,
N,
O,
Al,
Si,
P,
S,
Ca, Sc, Ti, V, Cr, Mn Fe, Co, Ni,
Zn, 
Ge, and
Sn,
were considered in the model-atmosphere calculations.
For all elements, we account for level dissolution (pressure ionization) following
\citet{hummermihalas1988} and \citet{hubenyetal1994}. 
Figure\,\ref{fig:HIoccu} demonstrates that our \ion{H}{i} model ion \sT{tab:modelatom}
includes all levels that are relevant in the line-forming region
$-4.5 < \log [m / (\mathrm{g/cm}^2)]$.
All model atmospheres presented here cover column densities $m$ of $-7.6 < \log m < 3.2$
\citep[cf\@.][]{beuermannetal2006}
represented by 90 depth points.

The model-atoms and respective absorption cross-sections for
Ca -- Ni were calculated via the recently registered VO service
TIRO\footnote{\url{http://astro.uni-tuebingen.de/~TIRO}}
that uses Kurucz's atomic data\footnote{\url{http://kurucz.harvard.edu/atoms.html}} 
and line lists \citep{kurucz1991, kurucz2009, kurucz2011}.

Ca, Sc, Ti, V, Cr, Mn, and Co lines are not identified. These were merged into a
generic model atom \citep{rauchdeetjen2003} with fixed solar abundance ratios. Then, we
performed test calculations and adjusted the abundance to a value 
($1.78 \times 10^{-6}$ by mass, the solar value is $9.93 \times 10^{-5}$) 
where all of its lines just fade in the noise of the 
observed spectra. All other model atoms were constructed from data retrieved from
the public T\"ubingen model-atom database
TMAD\footnote{\url{http://astro.uni-tuebingen.de/~TMAD}}.

In total, we considered 1038 atomic levels in NLTE combined with 4646 line transitions 
(for the number of individual iron-group lines, see Table\,\ref{tab:modelatom}) in the 
model-atmosphere calculations with 53\,203 frequency points within 
$1 \times 10^{12}\,\mathrm{Hz} \le \nu \le 3 \times 10^{17}\,\mathrm{Hz}$.
For the emergent spectra ($100\,\mathrm{\AA} \le \lambda \le 400\,000\,\mathrm{\AA}$,
686\,196 wavelength points),
we account for fine-structure splitting and used 1585 NLTE levels and
9721 respective line transitions.
The model-atom statistics are summarized in \ta{tab:modelatom}.
Figure\,\ref{fig:ionfrac} shows the ionization fractions of all elements in our final model.
It may be interesting to note that a single model atmosphere needs about one week to converge, i.e\@.
the absolute values of all relative corrections are below $10^{-4}$, 
on a 64\,bit, 2.66\,GHz compute core with 8\,GB memory.

\onltab{
\onecolumn
\begin{longtable}{rr@{~}lcccr}
\caption{\label{tab:modelatom}Statistics of our model atoms.
                              IG denotes a generic model atom consisting of Ca, Sc, Ti, V, Cr, Mn, and Co.
                              ``sample lines'' are individual Kurucz' lines that are sampled to
                              superlines for Ca - Ni \citep{rauchdeetjen2003}.}\\
\hline\hline
\noalign{\smallskip}
      & \multicolumn{2}{c}{ion~~~~}  &  NLTE levels &   LTE levels &  lines  & sample lines \\
\noalign{\smallskip}
\hline
\endfirsthead
\caption{continued.}\\
\hline\hline
\noalign{\smallskip}
      & \multicolumn{2}{c}{ion~~~~}  &  NLTE levels &   LTE levels &  lines  & sample lines \\
\noalign{\smallskip}
\hline
\noalign{\smallskip}
\endhead
\hline
\noalign{\smallskip}
\endfoot
\noalign{\smallskip}
      & H  & {\sc i}    &                       14 &            2 &     91 &               \\
      &    & {\sc ii}   &                        1 &          $-$ &    $-$ &               \\
\noalign{\smallskip}                                                              
      & He & {\sc i}    &                       29 &           74 &     69 &               \\
      &    & {\sc ii}   &                       16 &           16 &    120 &               \\
      &    & {\sc iii}  &                        1 &          $-$ &    $-$ &               \\
\noalign{\smallskip}                                                              
      & C  & {\sc ii}   &                        1 &           45 &      0 &               \\
      &    & {\sc iii}  &                       44 &           23 &    190 &               \\
      &    & {\sc iv}   &                       54 &            4 &    295 &               \\
      &    & {\sc v}    &                        1 &            0 &      0 &               \\
\noalign{\smallskip}                                                              
      & N  & {\sc ii}   &                        1 &          246 &      0 &               \\
      &    & {\sc iii}  &                       34 &           32 &    129 &               \\
      &    & {\sc iv}   &                       90 &            4 &    546 &               \\
      &    & {\sc v}    &                       54 &            8 &    297 &               \\
      &    & {\sc vi}   &                        1 &            0 &      0 &               \\
\noalign{\smallskip}                                                              
      & O  & {\sc ii}   &                        1 &           46 &      0 &               \\
      &    & {\sc iii}  &                       72 &            0 &    322 &               \\
      &    & {\sc iv}   &                       38 &           56 &    173 &               \\
      &    & {\sc v}    &                       76 &           50 &    472 &               \\
      &    & {\sc vi}   &                       54 &            8 &    291 &               \\
      &    & {\sc vii}  &                        1 &            0 &      0 &               \\
\noalign{\smallskip}                                                              
      & Al & {\sc ii}   &                        1 &            4 &      0 &               \\
      &    & {\sc iii}  &                        7 &           29 &     10 &               \\
      &    & {\sc iv}   &                        6 &          183 &      3 &               \\
      &    & {\sc v}    &                        6 &          223 &      4 &               \\
      &    & {\sc vi}   &                        1 &            0 &      0 &               \\
\noalign{\smallskip}                                                              
      & Si & {\sc iii}  &                       17 &           17 &     28 &               \\
      &    & {\sc iv}   &                       16 &            7 &     44 &               \\
      &    & {\sc v}    &                        1 &            0 &      0 &               \\
\noalign{\smallskip}                                                              
      & P  & {\sc iii}  &                        3 &            7 &      0 &               \\
      &    & {\sc iv}   &                       21 &           30 &      9 &               \\
      &    & {\sc v}    &                       18 &            7 &     12 &               \\
      &    & {\sc vi}   &                        1 &            0 &      0 &               \\
\noalign{\smallskip}                                                              
      & S  & {\sc iii}  &                        1 &          230 &      0 &               \\
      &    & {\sc iv}   &                       17 &           83 &     32 &               \\
      &    & {\sc v}    &                       39 &           71 &    107 &               \\
      &    & {\sc vi}   &                       25 &           12 &     48 &               \\
      &    & {\sc vii}  &                        1 &            0 &      0 &               \\
\noalign{\smallskip}                                        
      & Fe & {\sc iii}  &                        7 &            0 &     25 &     537\,689  \\
      &    & {\sc iv}   &                        7 &            0 &     25 &  3\,102\,371  \\
      &    & {\sc v}    &                        7 &            0 &     25 &  3\,266\,247  \\
      &    & {\sc vi}   &                        7 &            0 &     33 &     991\,935  \\
      &    & {\sc vii}  &                        7 &            0 &     39 &     200\,455  \\
      &    & {\sc viii}  &                       1 &            0 &      0 &            0  \\
\noalign{\smallskip}                                              
      & Ni & {\sc iii}  &                        7 &            0 &     22 &  1\,033\,920  \\
      &    & {\sc iv}   &                        7 &            0 &     25 &  2\,512\,561  \\
      &    & {\sc v}    &                        7 &            0 &     27 &  2\,766\,664  \\
      &    & {\sc vi}   &                        7 &            0 &     27 &  7\,408\,657  \\
      &    & {\sc vii}  &                        7 &            0 &     33 &  4\,195\,381  \\
      &    & {\sc viii} &                        1 &            0 &      0 &            0  \\
\noalign{\smallskip}                                              
      & IG & {\sc iii}  &                        1 &            0 &      0&             0  \\
      &    & {\sc iv}   &                        7 &            0 &     25 &  1\,579\,918  \\
      &    & {\sc v}    &                        7 &            0 &     23 &  2\,230\,921  \\
      &    & {\sc vi}   &                        7 &            0 &     25 &  1\,455\,521  \\
      &    & {\sc vii}  &                        7 &            0 &     24 &  1\,129\,512  \\
      &    & {\sc viii} &                        1 &            0 &      0 &            0  \\
\noalign{\smallskip}                                        
      & Zn & {\sc ii}   &                        1 &            5 &      0 &               \\
      &    & {\sc iii}  &                        2 &           10 &      0 &               \\
      &    & {\sc iv}   &                       31 &            0 &     87 &               \\
      &    & {\sc v}    &                        5 &           15 &      0 &               \\
      &    & {\sc vi}   &                        1 &            0 &      0 &               \\
\noalign{\smallskip}                                        
      & Ge & {\sc iii}  &                        1 &           15 &      0 &               \\
      &    & {\sc iv}   &                        8 &            1 &      8 &               \\
      &    & {\sc v}    &                       85 &            0 &    878 &               \\
      &    & {\sc vi}   &                       11 &           25 &      0 &               \\
      &    & {\sc vii}  &                        1 &            0 &      0 &               \\
\noalign{\smallskip}                                                         
      & Sn & {\sc iii}  &                        3 &           18 &      2 &               \\
      &    & {\sc iv}   &                        6 &            4 &      1 &               \\
      &    & {\sc v}    &                        5 &            4 &      0 &               \\
      &    & {\sc vi}   &                        6 &            0 &      0 &               \\
      &    & {\sc vii}  &                        1 &            0 &        &               \\
\hline           \hline
\noalign{\smallskip}                                        
total &    &         70 &                     1038 &         1614 &   4646 & 32\,411\,752  \\
\hline
\end{longtable}
\twocolumn
}

\onlfig{
\begin{figure*}
  \resizebox{\hsize}{!}{\includegraphics{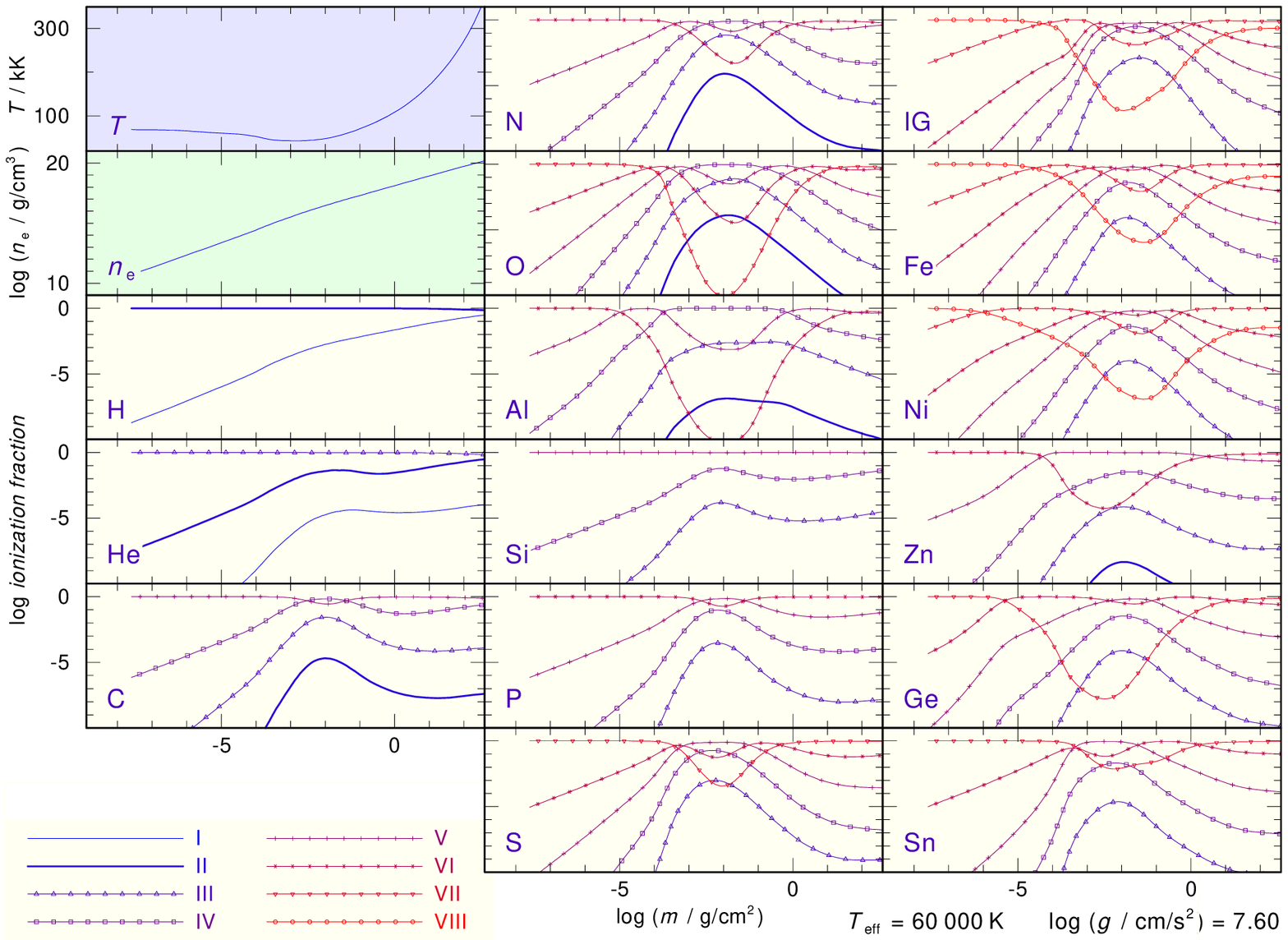}}
  \caption{Temperature and density structure and ionizations fractions of our model
           with \Teffw{60\,000} and \loggw{7.60}.
           IG denotes a generic model atom consisting of Ca, Sc, Ti, V, Cr, Mn, and Co.} 
  \label{fig:ionfrac}
\end{figure*}
}

For the calculation of synthetic \ion{H}{i} line profiles, we use Stark line-broadening
tables provided by \citet{tremblaybergeron2009}. 
For those lines, where no broadening tables are available, TMAP uses an approximate
formula, as described in \citet[Eqs\@. 1 - 5]{ziegleretal2012}.

We started with a model with \Teffw{60\,920} and \loggw{7.55}
\citep[the values of][]{gianninasetal2011} and the element abundances from Table\,\ref{tab:prevabund}.
Next, we adjusted these abundances to best reproduce the respective spectral lines.
We then calculated an extended grid of 234 model atmospheres
($48\,000\,\mathrm{K} \le \Teff \le 68\,000\,\mathrm{K}$ in steps of $\le 1000\,\mathrm{K}$ and
$7.35 \le \logg \le 7.75$ in steps of $0.05$ (some of the hotter models are calculated only for
$\logg \le 7.60$). For this grid, we extensively used compute resources of the 
bwGRiD\footnote{\url{http://www.bw-grid.de/en/the-bwgrid/}} in addition. Although this highly speeded up the model-grid
calculation, the wide parameter range and the large number (15) of parameters to adjust simultaneously
did not allow us to take a statistical approach in the spectral analysis
\citep[$\chi^2$ method like e.g\@. in][]{gianninasetal2011} on a reasonable time scale. We therefore
need to rely upon our ``$\chi$-by-eye'' methods. All SEDs that were calculated for this analysis are available via
TheoSSA\footnote{\url{http://dc.g-vo.org/theossa}}.

In a first analysis step, we will determine \logg based on UV and optical observations.
Then, we will determine \Teff precisely based on ionization equilibria of the metals which are
sensitive indicators.
Subsequently, we will adjust the abundances again and verify our \Teff and \logg results.

\subsection{Surface gravity and effective temperature}
\label{sect:tefflogg}

\begin{figure*}
  \resizebox{\hsize}{!}{\includegraphics{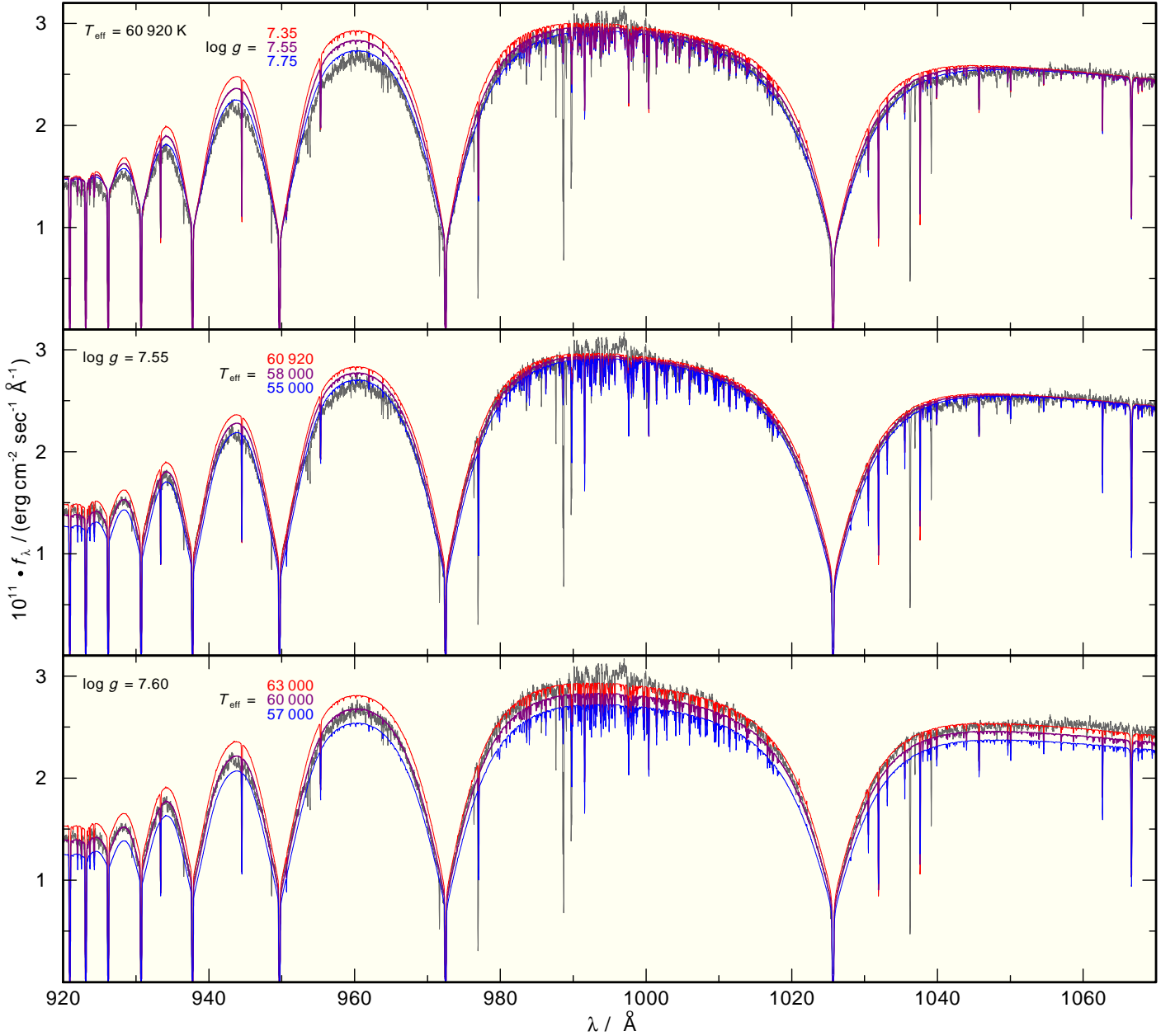}}
  \caption{Section of the FUSE observation compared with our model fluxes with different \Teff and \logg.
           In the top and middle panels, the synthetic fluxes are normalized to the observed flux at 1000\,\AA\ and
           in the bottom panel to the observed K magnitude (see Fig.\,\ref{fig:ebv}). $E_{B-V}$ and $N_\ion{H}{i}$
           are applied using our results from Sect.\,\ref{sect:ism}.
          } 
  \label{fig:fuv}
\end{figure*}

The dependency of the synthetic flux level on \Teff and \logg for fixed abundances is
demonstrated in Fig.\,\ref{fig:fuv}, where we compare the observed and synthetic fluxes in the FUV.
In the top panel, it is obvious that at a constant \Teffw{60\,920}, a \logg higher by 0.2\,dex than 
\loggw{7.55} measured by \citet{gianninasetal2011} is necessary to reproduce the Lyman-line decrement. For
a fixed \loggw{7.55} (middle panel), a lower \Teff ($\Delta$\Teffw{6000}) improves the agreement between model and
observation. The bottom panel shows that at values within the (statistical) error ranges from the
\ion{H}{i} Balmer-line analysis, \Teffw{60\,000} and \loggw{7.60}
\citep[cf\@. Table\,\ref{tab:previous}][]{gianninasetal2011}, a good agreement for both, line profiles and
decrement, is achieved.

\begin{figure}
  \resizebox{\hsize}{!}{\includegraphics{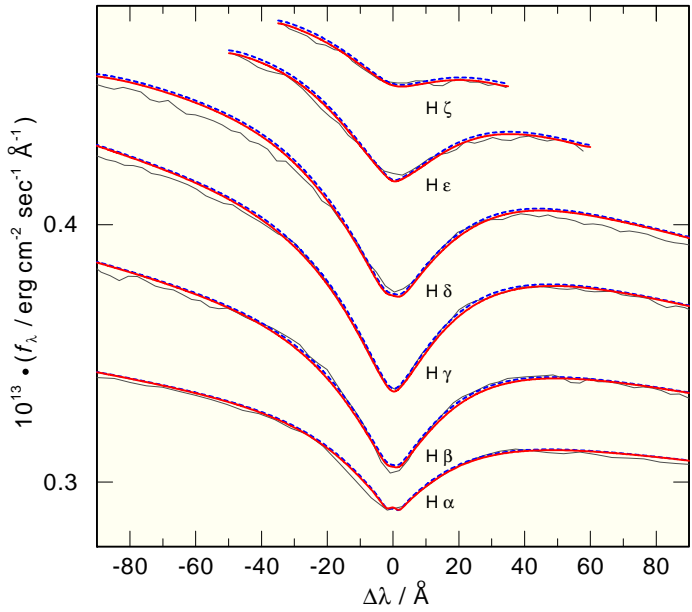}}
  \caption{Synthetic \ion{H}{i} Balmer-line profiles compared with the STIS LR observation.
           The observation is shifted by factors of 0.921, 0.323, 0.232, 0.202, 0.193, and
           0.188 for H\,$\alpha$ - H\,$\zeta$, respectively, to fit into this plot.
           Red, full line: \Teffw{60\,000} and \loggw{7.60}; 
             blue, dashed: \Teffw{60\,920} and \loggw{7.55}.
          } 
  \label{fig:optical}
\end{figure}

\begin{figure}
  \resizebox{\hsize}{!}{\includegraphics{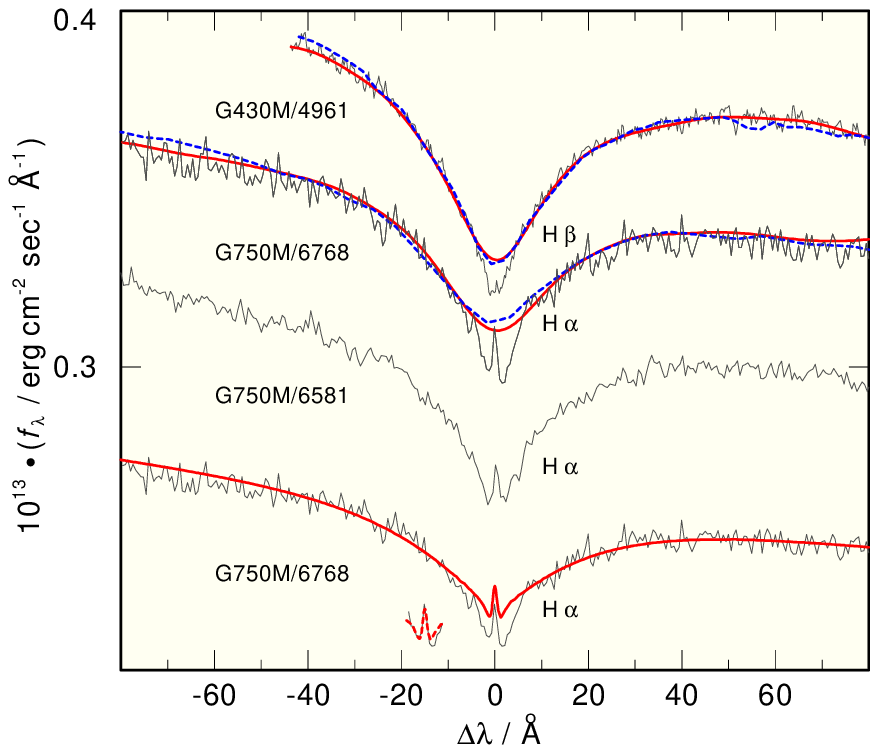}}
  \caption{STIS H\,$\alpha$ and H\,$\beta$ low-resolution (dashed, blue) and
           medium-resolution (gray lines) observations (labeled with medium-resolution
           grating/central wavelength in \AA). Because of uncertainty in the flux
           calibration, the medium-resolution data are normalized to the low-resolution
           flux. The flux around H\,$\beta$ in the top plot is multiplied by a factor of 0.35.
           The red lines in the upper two plots are the medium-resolution ($R \approx 6000$) spectra
           degraded to the low resolution ($R \approx 500$) and agree with the low resolution (blue
           dash) within the uncertainty of the $R=500$ resolution. The lower two plots are
           shifted down by $0.035$ and $0.07 \times  10^{-13}$ flux units, respectively. In the lowest
           plot, the model is overplotted in red after smoothing to the medium resolution.
           While the model H\,$\alpha$ absorption is somewhat too weak, the central emission
           agrees with the observation within the uncertainty of the resolution (insert).
          } 
  \label{fig:hrlr}
\end{figure}

This was not expected from the outset although \citet{barstowetal1998} found a relatively good agreement of
\Teff and \logg from \ion{H}{i} Lyman and Balmer lines in the heavy-metal rich models (Table\,\ref{tab:previous}).
The later analysis by \citet[][Table\,\ref{tab:previous}]{barstowetal2001} shows strong deviations in \logg
between optical and FUV analyses. Figure\,\ref{fig:optical} shows a comparison of 
synthetic \ion{H}{i} Balmer line profiles with optical observations. The deviation between the \Teffw{60\,920} / \loggw{7.55} and
the 60\,000 / 7.60 ones is minor. 

In addition to the low-resolution ($R \approx 500$) optical spectrum, medium-resolution 
($R \approx 6000$) observations of H\,$\alpha$ and H\,$\beta$ are shown is
Fig.\,\ref{fig:hrlr}. The agreement among the STIS low and medium resolution is excellent.
While the model absorption is a bit weak as shown in the lower plot of Fig.\,\ref{fig:hrlr}, the central NLTE emission
reversal agrees well with the observations.

Although we cannot reproduce H\,$\alpha$ and H\,$\beta$ in detail in the medium-resolution spectrum,
this has no significant influence on our determination of \Teff and \logg because the higher members
of the \ion{H}{i} Balmer series form much deeper in the atmosphere where the influence of 
metal opacities is less important \citep[cf\@.][]{napiwotzkirauch1994}.
We adopt \loggw{7.60}.

In the next step of this analysis, 
we evaluate ionization equilibria of metals that exhibit lines of successive
ionization stages. 
Figures\,\ref{fig:teff}, \ref{fig:teffFe}, and \ref{fig:teffNi} show some strategic lines for 
this \Teff determination. In total, we can use eight elements and lines of
\ion{C}{iii}  - \ion{}{iv} ,
\ion{N}{iii}  - \ion{}{v},
\ion{Si}{iii} - \ion{}{iv},
\ion{P}{iv}   - \ion{}{v},
\ion{S}{iv}   - \ion{}{v}, 
\ion{Fe}{iv}   - \ion{}{vi}, 
\ion{Ni}{iv}   - \ion{}{vi}, and 
\ion{Ge}{iv}  - \ion{}{v}.
\Teffw{60\,000 \pm 2000} reproduces well all these equilibria simultaneously.
For our further analysis, we adopt \Teffw{60\,000}.

\begin{landscape}
\addtolength{\textwidth}{6.3cm} 
\addtolength{\evensidemargin}{0cm}
\addtolength{\oddsidemargin}{0cm}
\begin{figure*}[ht]
  \includegraphics[trim=0 0 0 0,height=14.1cm,angle=0]{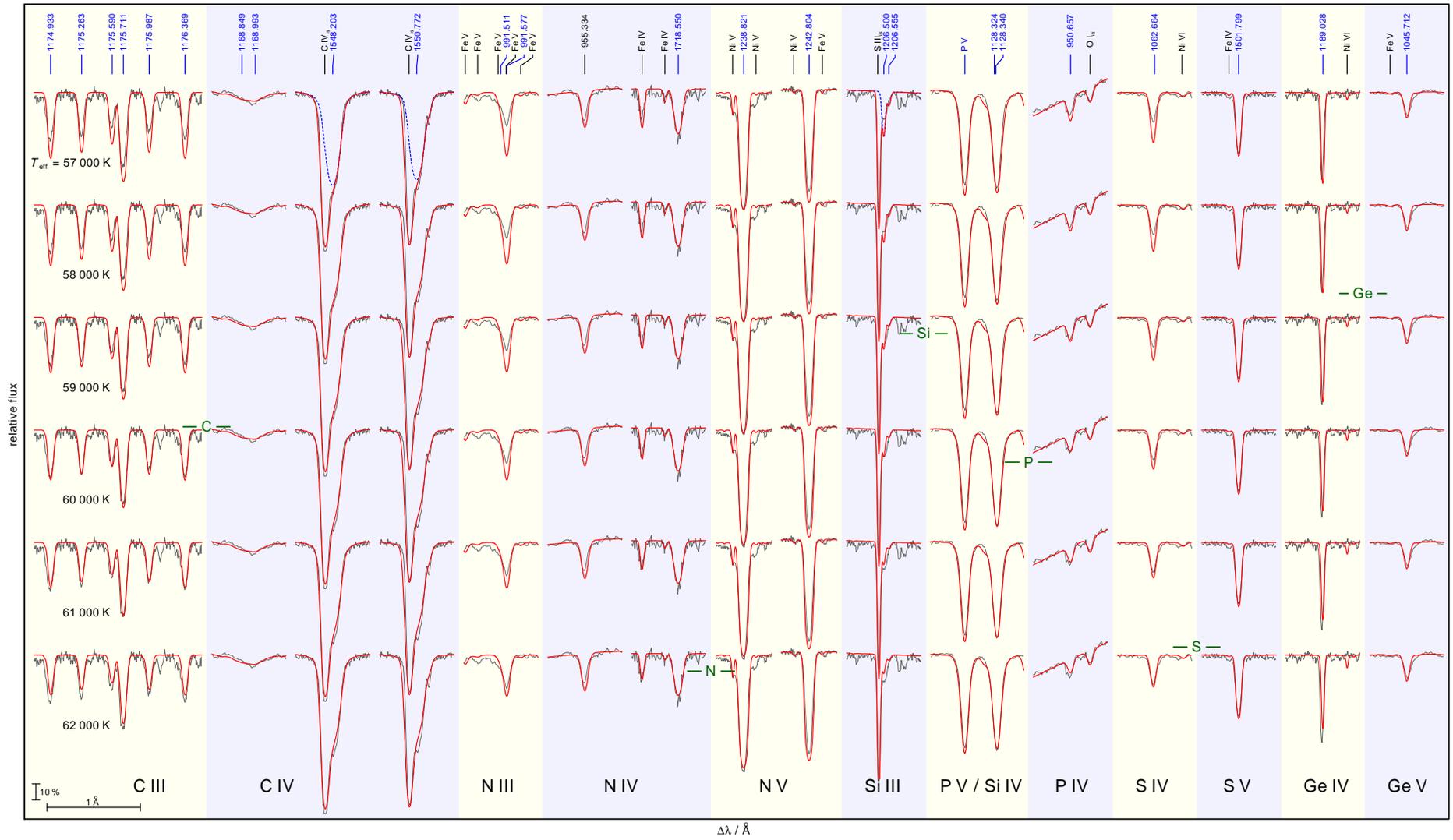}
  \caption{Synthetic lines profiles of
           \ion{C}{iii}  - \ion{}{iv} ,
           \ion{N}{iii}  - \ion{}{v},
           \ion{Si}{iii} - \ion{}{iv},
           \ion{P}{iv}   - \ion{}{v},
           \ion{S}{iv}   - \ion{}{v}, and 
           \ion{Ge}{iv}  - \ion{}{v} calculated from models with
           $57\,000\,\mathrm{K} \le T_\mathrm{eff} \le 62\,000\,\mathrm{K}$ and \loggw{7.6}
           compared with FUSE and STIS observations. 
           The small horizontal bars, labeled with the element's name, indicate \Teff
           where the ionization balance is best reproduced.
           The lines are identified at top. 
           ``is'' denotes interstellar origin. 
           In the cases of 
           \ion{C}{iv} $\lambda\lambda\, 1548, 1550$\,\AA, 
           \ion{C}{iv} $\lambda\lambda\, 1548, 1550$\,\AA, and
           \ion{Si}{iii} $\lambda\lambda\, 1206.500, 1206.555$\,\AA,
           the photospheric line profile is shown by a dashed, blue 
           line (for the \Teffw{57\,000} model at top only).
          }
  \label{fig:teff}
\end{figure*}
\end{landscape}

\onlfig{
\begin{figure*}
  \resizebox{\hsize}{!}{\includegraphics{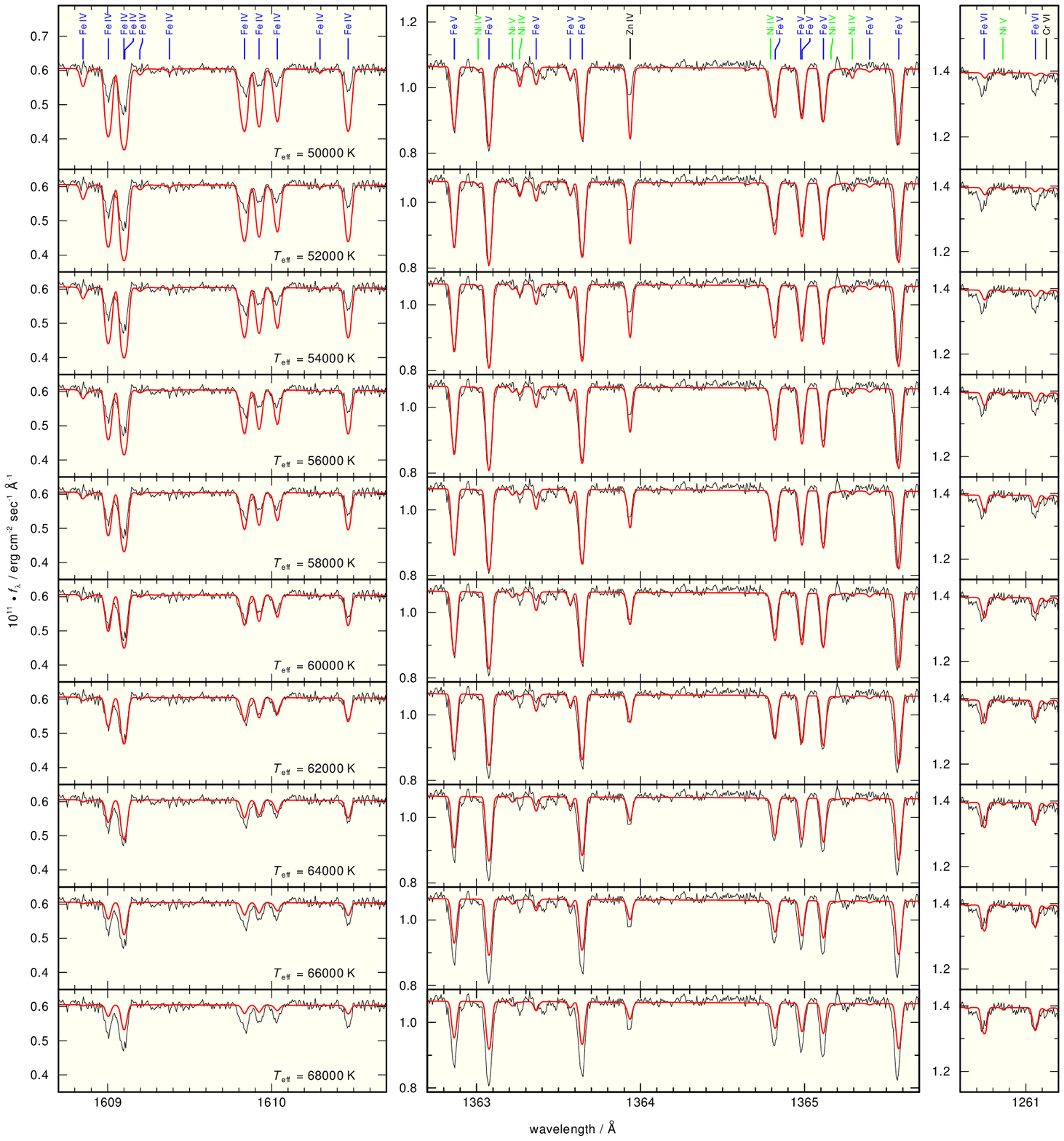}}
  \caption{Same as Fig.\,\ref{fig:teff} for \ion{Fe}{iv} - \ion{}{vi} lines (from left to right panels, marked blue in the top panels) only.} 
  \label{fig:teffFe}
\end{figure*}
}

\onlfig{
\begin{figure*}
  \resizebox{\hsize}{!}{\includegraphics{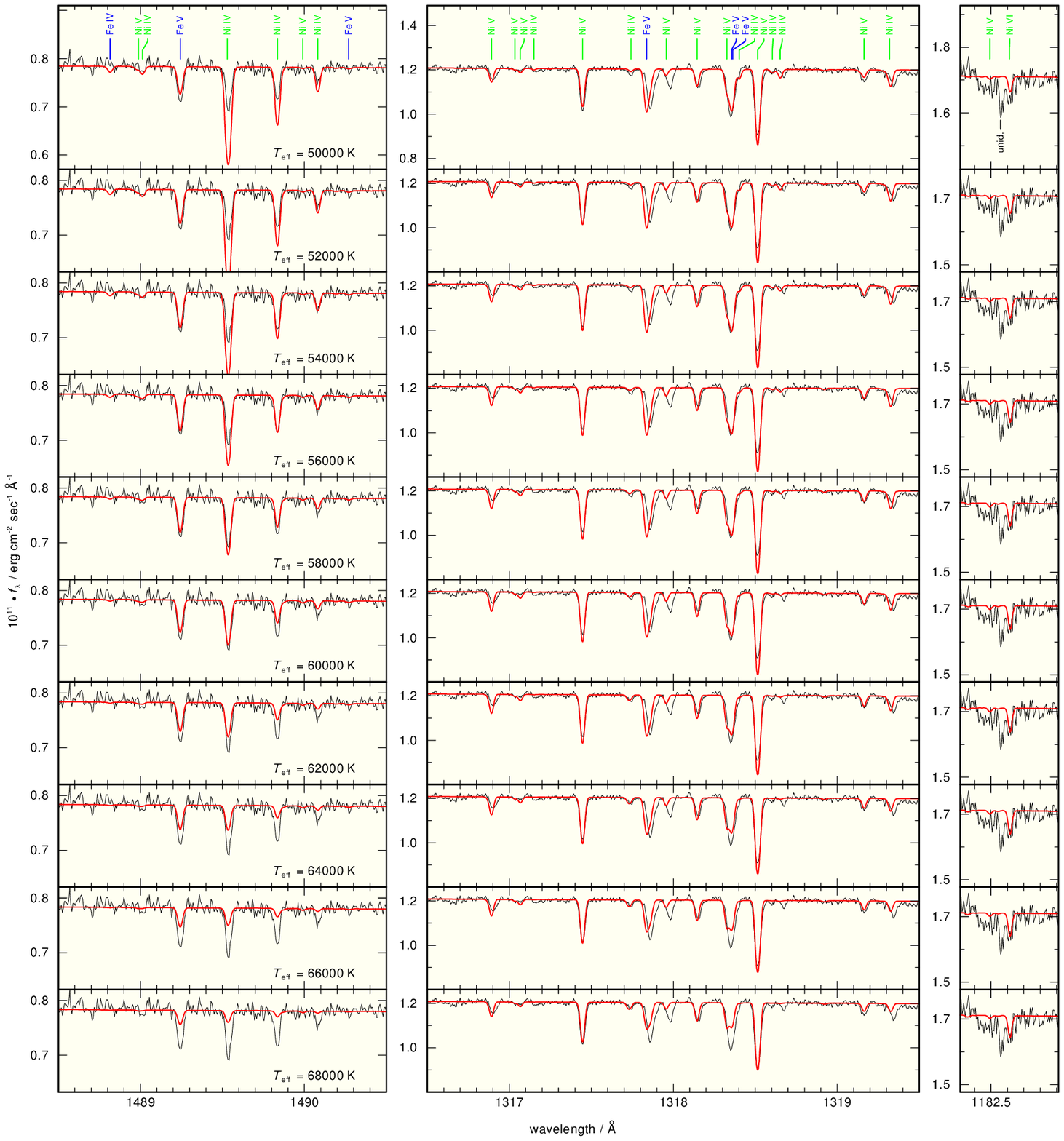}}
  \caption{Same as Fig.\,\ref{fig:teff} for \ion{Ni}{iv} - \ion{}{vi} lines (from left to right panels, marked green in the top panels) only.} 
  \label{fig:teffNi}
\end{figure*}
}

\subsection{Photospheric abundances}
\label{sect:abund}

In the following, we use logarithmic mass fractions for all abundance values, if not otherwise
mentioned.
Previously determined abundances and respective references are summarized in Table\,\ref{tab:prevabund}.
In the following, we will briefly mention the strategic lines for the abundance 
determinations and note abnormalities for an element selection only. Most of the identified 
metals exhibit lines of at least two subsequent ionization stages and some of these lines 
were already used for the determination of \Teff (Sect.\,\ref{sect:tefflogg}). The abundances 
were then adjusted to achieve best line fits.
Two large plots (German DIN\,A0 size) are provided in the online material that show a comparison of our final model with the 
observation in the FUSE and STIS wavelength ranges (in total $911 - 1750$\,\AA). 
They include all line identifications (FUSE/STIS wavelength range), e.g\@. 
  2/421 \ion{Fe}{iv}, 
144/815 \ion{Fe}{v}, 
  1/52  \ion{Fe}{vi}, 
  1/236 \ion{Ni}{iv}, 
 13/690 \ion{Ni}{v}, and
  9/43 \ion{Ni}{vi} lines. 
These numbers are much higher than those of \citet[][106 \ion{Fe}{v} and 44 \ion{Ni}{v} lines in the STIS wavelength range]{prevaletal2013}.
The recent work of \citet{berengutetal2013} to employ \gb as a stellar laboratory to determine the
fine-structure constant is based on the latter list and may, thus, not fully exploit capacity of all the 
available STIS spectra of \gb.

Our line identifications are also summarized in Table\,\ref{tab:lineids}, whereas
Table\,\ref{tab:unid} gives a list of the strongest unidentified lines.

\onltab{
\onecolumn
\begin{longtable}{r@{.}lr@{\,}lr@{.}lccr@{.}lr@{\,}llr@{.}lc}
\caption{\label{tab:lineids}Identified lines. 
                            $\lambda_\mathrm{rest}$ is the wavelength in the stellar rest frame as given by Kurucz (iron-group elements)
                            or NIST.
                            $\lambda_\mathrm{theo}$ is the theoretical wavelength of lines with ISM origin.
                            The V, Cr, Mn, and Co line identifications are uncertain.} \\
\hline\hline
\noalign{\smallskip}
   \multicolumn{2}{c}{$\lambda_\mathrm{rest}$ / \AA} & \multicolumn{2}{c}{ion} & \multicolumn{2}{c}{$\lambda_\mathrm{theo}$ / \AA} & comment & ~~~ &
   \multicolumn{2}{c}{$\lambda_\mathrm{rest}$ / \AA} & \multicolumn{2}{c}{ion} & \multicolumn{2}{c}{$\lambda_\mathrm{theo}$ / \AA} & comment       \\
\noalign{\smallskip}
\hline
\endfirsthead
\caption{continued.}\\
\hline\hline
\noalign{\smallskip}
  \multicolumn{2}{c}{$\lambda_\mathrm{rest}$ / \AA} & \multicolumn{2}{c}{ion} & \multicolumn{2}{c}{$\lambda_\mathrm{theo}$ / \AA} & comment & ~~~ &
  \multicolumn{2}{c}{$\lambda_\mathrm{rest}$ / \AA} & \multicolumn{2}{c}{ion} & \multicolumn{2}{c}{$\lambda_\mathrm{theo}$ / \AA} & comment       \\
\noalign{\smallskip}
\hline
\noalign{\smallskip}
\endhead
\hline
\noalign{\smallskip}
\endfoot
\noalign{\smallskip}
 913&173 & H&{\sc i}    &              913&215 & ISM && 1347&210 & Ni&{\sc v}   & \multicolumn{2}{c}{} &     \\
 913&298 & H&{\sc i}    &              913&339 & ISM && 1347&720 & Ni&{\sc v}   & \multicolumn{2}{c}{} &     \\
 913&439 & H&{\sc i}    &              913&480 & ISM && 1347&952 & Zn&{\sc iv}  & \multicolumn{2}{c}{} &     \\
 913&600 & H&{\sc i}    &              913&641 & ISM && 1348&843 & Fe&{\sc v}   & \multicolumn{2}{c}{} &     \\
 913&784 & H&{\sc i}    &              913&826 & ISM && 1349&154 & Ni&{\sc iv}  & \multicolumn{2}{c}{} &     \\
 913&997 & H&{\sc i}    &              914&039 & ISM && 1349&335 & Ni&{\sc v}   & \multicolumn{2}{c}{} &     \\
 914&245 & H&{\sc i}    &              914&286 & ISM && 1349&783 & Ni&{\sc v}   & \multicolumn{2}{c}{} &     \\
 914&535 & H&{\sc i}    &              914&576 & ISM && 1349&874 & Zn&{\sc iv}  & \multicolumn{2}{c}{} &     \\
\hline
\end{longtable}
\twocolumn
}

\begin{table}\centering
\caption{Wavelengths (in \AA) of unidentified strong ($W_\lambda > 10\,\mathrm{m\AA}$), likely photospheric lines in the FUSE and STIS observations.}
\label{tab:unid}
\begin{tabular}{r@{.}lr@{.}lr@{.}lr@{.}lr@{.}l}
\hline
\noalign{\smallskip}
\multicolumn{2}{l}{FUSE}  & \multicolumn{8}{c}{} \\                             
  989&11 & 1029&44 & 1133&04 & 1142&91 & \multicolumn{2}{l}{} \\
\hline
\noalign{\smallskip}
\multicolumn{2}{l}{STIS}  & \multicolumn{8}{c}{} \\                             
1157&60 & 1173&29 & 1183&82 & 1201&51 & 1283&52 \\ 
1158&08 & 1174&34 & 1186&08 & 1201&81 & 1306&04 \\ 
1158&76 & 1176&11 & 1186&27 & 1202&43 & 1331&19 \\ 
1165&31 & 1176&52 & 1186&59 & 1204&47 & 1385&32 \\ 
1166&80 & 1176&64 & 1187&70 & 1227&55 & 1389&89 \\ 
1171&12 & 1176&87 & 1190&15 & 1253&67 & 1398&21 \\ 
1171&19 & 1176&98 & 1192&01 & 1253&87 & 1411&46 \\ 
1171&45 & 1177&06 & 1194&16 & 1258&81 & 1516&68 \\ 
1172&18 & 1178&68 & 1198&15 & 1260&48 & 1520&64 \\ 
1172&34 & 1182&00 & 1201&30 & 1272&98 & 1525&32 \\ 
1173&22 & 1183&37 & 1201&47 & 1281&37 & 1538&94 \\ 
\hline
\end{tabular}
\end{table}

\subsubsection{Helium}
\label{sect:helium}

\begin{figure}
  \resizebox{\hsize}{!}{\includegraphics{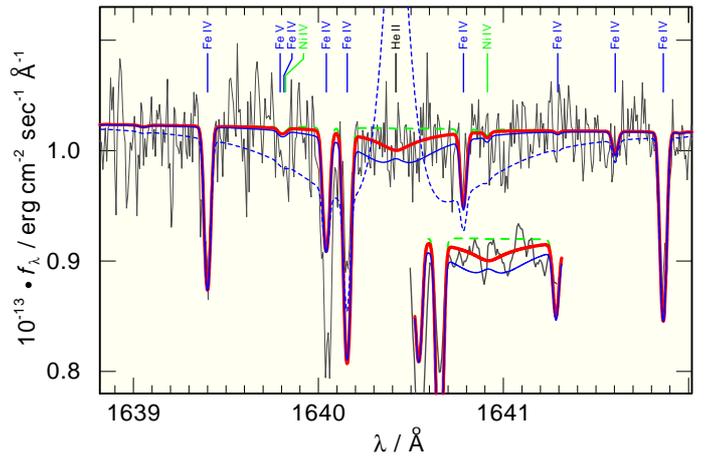}}
  \caption{Synthetic spectrum around \ion{He}{ii} $\lambda\, 1640.42$\,\AA\ compared with the STIS observation.
           the He abundances are
           green, dashed: $-10.0$,
           red, thick: $-4.7$,
           blue, thin: $-4.2$, 
           blue, dashed: $-3.1$.
           The insert shows the region $\Delta\lambda = \pm 0.4$\,\AA\ around the \ion{He}{ii} line.
           For comparison, the observation was smoothed with a low-pass filter \citep[$n=15$, $m=4$]{savitzkygolay1964}.}
  \label{fig:he}
\end{figure}

The first analyses revealed only upper limits for the He abundance, e.g\@.
He $< -3.1$ and $< 4.1$ \citep[respectively]{vennesetal1996,gundersonetal2001}. 
\citet{cruddaceetal2002} determined He\,$ = -4.2 \pm 0.1$ using high-resolution
EUV spectroscopy. An attempt to identify and 
measure \ion{He}{ii} Lyman lines ($n$ - $n'$ = 1 - 4, 1 - 5) with
J-PEX\footnote{Joint Astrophysical Plasmadynamic Experiment} \citep{barstowetal2005} was not successful.
Our models show that \Ionw{He}{2}{1640} (2 - 3) should be clearly visible at He\,$= -3.7$ and $-4.2$ and
disappears in the noise of the observation only at about He\,$ < -4.7$ (Fig.\,\ref{fig:he}).
We adopt this upper-limit value for our models.

\subsubsection{Carbon, nitrogen, and oxygen}
\label{sect:cno}

\ion{C}{iii} and \ion{C}{iv} lines are visible in the observation.
\Ionw{C}{3}{977.02} and \Ionww{C}{4}{1548.20, 1550.77} have strong ISM blends. In case of
the latter, the photospheric component can be separated and modeled (Fig.\,\ref{fig:teff}). 
At C\,$ = -5.15$, lines of both ions are well reproduced.

\ion{N}{iii} - \ion{}{v} lines are found in the observation, they are all well matched at
N\,$ = -5.58$  (Fig.\,\ref{fig:teff}).

\begin{figure}
  \resizebox{\hsize}{!}{\includegraphics{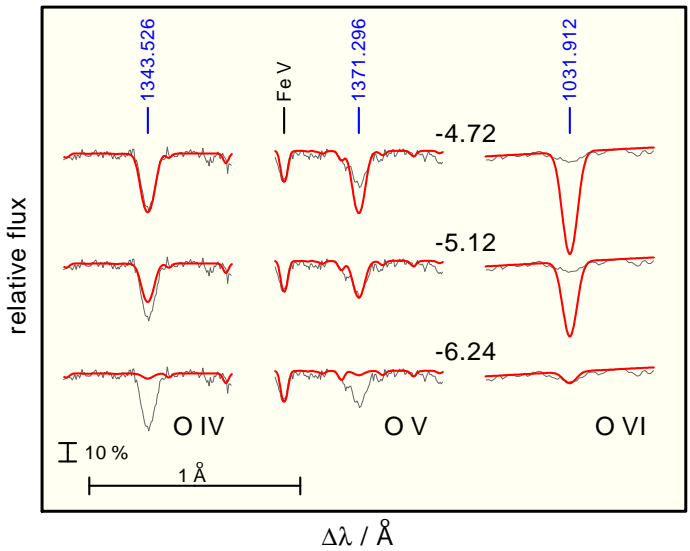}}
  \caption{The most prominent \ion{O}{iv}, \ion{O}{v}, and \ion{O}{vi} lines in the STIS and
           FUSE observations compared with our theoretical line profiles calculated with three
           O abundances (O\,$ = -4.72, -5.12, -6.24$, from top to bottom). 
           } 
  \label{fig:ostat}
\end{figure}

\citet{vennesetal2000} encountered deviations between oxygen abundances determined
from \ion{O}{iv} and \ion{O}{v} lines in an analysis of \object{Feige\,24}. The
\ion{O}{v} abundance was 0.5\,dex higher in their LTE model approach. In their NLTE
models, they found that the \ion{O}{iv} / \ion{O}{v} ionization equilibrium was
overcorrected by $-0.8$\,dex. They suggested an inhomogeneous stratification of
O in the atmosphere. \citet{venneslanz2001} discovered that a similar problem exists
in \gb, with an overcorrection of $-0.6$\,dex. Consequently they assumed that in both stars,
the interplay between selective radiation pressure and gravity in diffusive equilibrium
are the key processes for this phenomenon.
Fig.\,\ref{fig:ostat} shows the same deviation in our models. While 
\Ionww{O}{4}{1338.615, 1342.990, 1343.526} are well fitted at O\,$ = -4.72$,
\Ionw{O}{5}{1371.296} is apparently much stronger than observed. It is matched with an O
abundance that is reduced by $-0.4$\,dex.

In the FUSE observation, only the short wavelength component of the 
\Ionww{O}{6}{1031.912, 1037.614} resonance doublet is detectable.
The unexpected weakness of this doublet was already reported by \citet{oegerleetal2005}. 
\citet{dickinsonetal2012} verified that it stems from the photosphere.
The \ion{O}{vi} resonance doublet in our models is is even stronger, 
compared to \ion{O}{iv} and \ion{O}{v} lines, requiring a reduction of
the O abundance by about -1.5\,dex (Fig\,\ref{fig:ostat}).
\citet{dickinsonbarstowhubeny2012} encountered a similar problem with
enigmatically deep line profiles of the \ion{N}{v} resonance doublet 
in their models.
We revisit the problem with the oxygen abundances derived from different
ionization stages in Sections\,\ref{sect:profile} and \ref{sect:diffusion}
in detail.

\subsubsection{Aluminum, silicon, phosphorus, and sulfur}
\label{sect:alsips}

\begin{figure}
  \resizebox{\hsize}{!}{\includegraphics{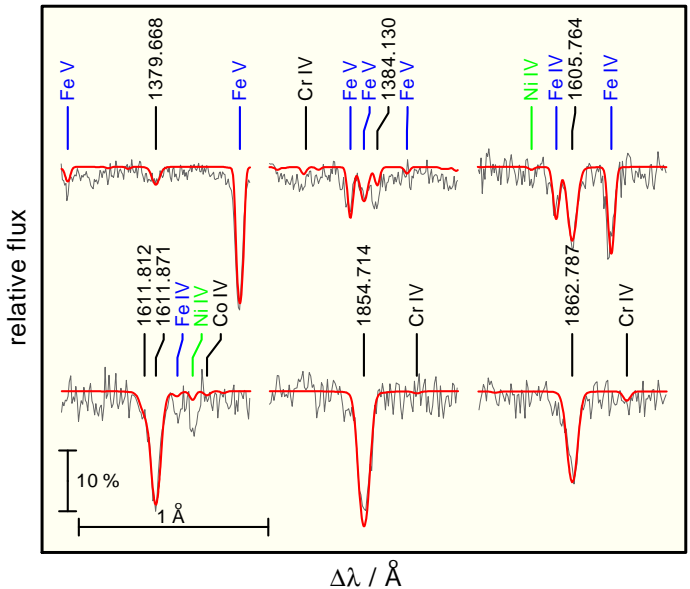}}
  \caption{Theoretical \ion{Al}{iii} line profiles calculated from our final model
           compared with the STIS observation.
           } 
  \label{fig:alion}
\end{figure}

\citet{holbergetal1998} identified the \Ionww{Al}{3}{1854.72, 1862.79}
resonance doublet in the IUE NEWSIPS SWP Echelle Data Set\footnote{\url{http://vega.lpl.arizona.edu/newsips/}},
and \citet{holbergetal2003} measured Al\,$ = -5.08$.
We could newly identify some other \ion{Al}{iii} lines.
We derive Al\,$ = -4.95$, well in agreement with the \citet{holbergetal2003} value (Fig.\,\ref{fig:alion}).

\ion{Si}{iii} - \ion{}{iv},
\ion{P}{iv} - \ion{}{v}, and
\ion{S}{v} - \ion{}{vi}
lines are identified.
We determine 
Si\,$ = -4.30$,
 P\,$ = -5.81$, and
 S\,$ = -5.24$ (Fig.\,\ref{fig:teff}).

\subsubsection{Iron-group elements}
\label{sect:irongroup}

Many hundreds of lines of \ion{Fe}{iv} - \ion{}{vi} and \ion{Ni}{iv} - {}{vi} are identified
(Table\,\ref{tab:lineids}).
They are best reproduced at Fe\,$ = -3.30$ and Ni\,$ = -4.45$.
Note that the Ni/Fe abundance ratio is about 25\,\% higher than the solar ratio.
Some of these lines are shown in various figures in this paper, please have a look at the
two large online figures that show the complete FUSE and STIS wavelength ranges.
An animation of STIS wavelength range can been seen at 
\url{http://astro.uni-tuebingen.de/~rauch/A0_E140H_SW.gif} as well.

In Fig.\,\ref{fig:alion} 
three lines of \ion{Cr}{iv} and 
one of \ion{Co}{iv} 
are visible in the synthetic spectrum of our final
model. These are weak and comparable to the noise of the observation. Although one may 
be tempted to believe the presence of \Ionw{Cr}{4}{1863.075}, we take this as a hint that a log mass fraction 
of $-5.75$ for the generic model atom is reasonable and adopt this as an upper limit for our analysis. 
This is, within the error limits,
in agreement with the upper limits for Cr, Mn, and Co of about $-6.2$ that was found by 
\citet{holbergetal2003}.

\citet{prevaletal2013} suggested that the unidentified line at 1272.98\,\AA\ is a \ion{V}{iv} line. 
Since many other \ion{V}{iv} lines with much stronger $\log gf$ values 
($g$ is the statistical weight of the lower atomic level and 
 $f$ is the oscillator strength of the line transition)
from Kurucz's POS line lists
(with good wavelengths) are not present in the spectrum, e.g\@. \Ionww{V}{4}{1355.127, 1419.577, 1426.647}
(all more than ten times higher $\log gf$) therefore this identification appears to be very unlikely.

We mention here that we find deviations between Kurucz's POS wavelengths and the observation of up to
0.05\,\AA. In addition, Fig.\,\ref{fig:he} shows that the strengths of \Ionw{Fe}{4}{1640.042} and \Ionw{Fe}{4}{1640.155}
in the model are the opposite way around in the observation.

\begin{figure}
  \resizebox{\hsize}{!}{\includegraphics{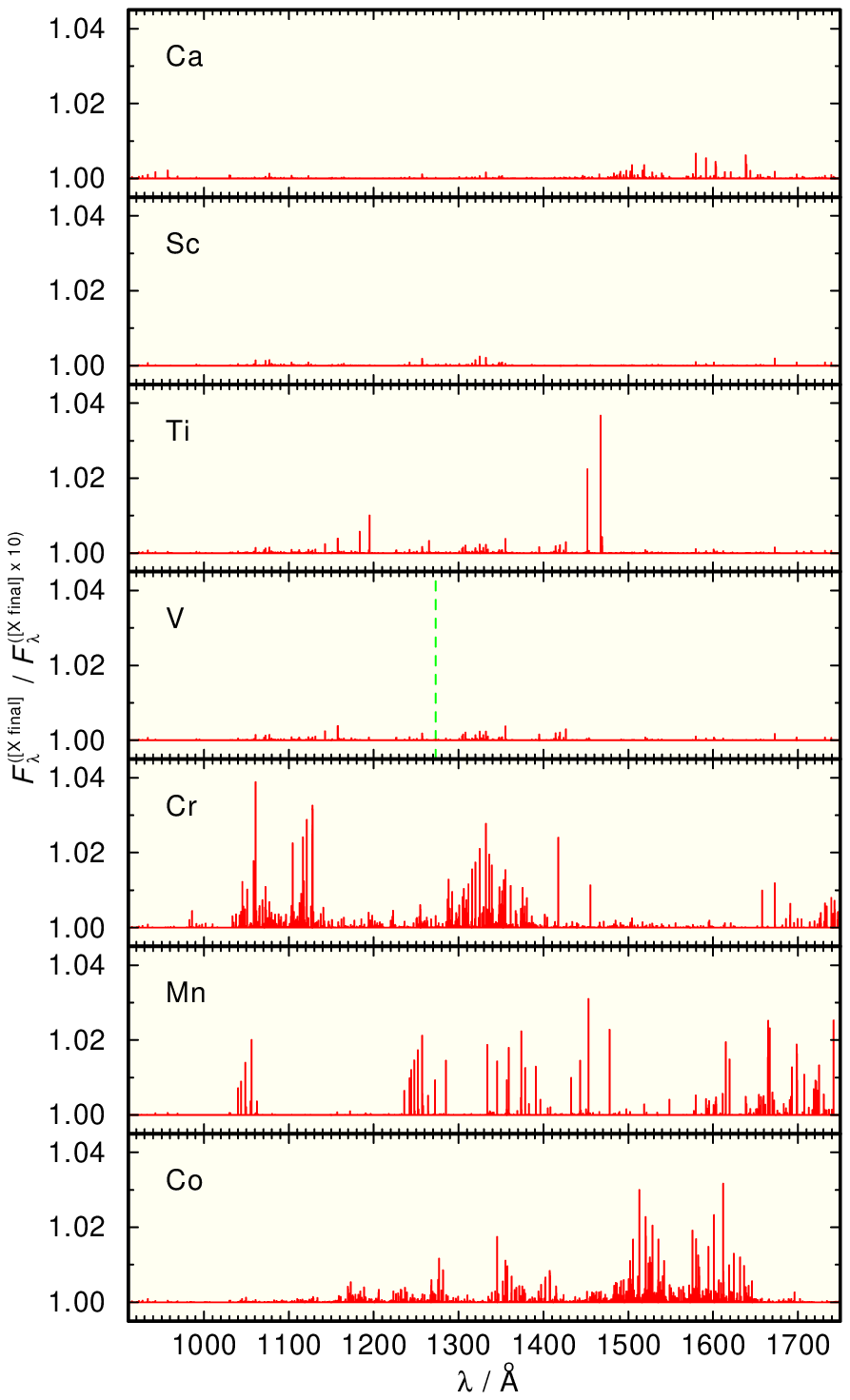}}
  \caption{Flux ratio of our final model and a model with ten times increased abundance of element X
           (Ca to Co, from top to bottom). 
           The location of \ion{V}{iv} $\lambda 1272.98$\,\AA\ is marked.
          } 
  \label{fig:geninc}
\end{figure}

Figure\,\ref{fig:geninc} shows a comparison of models (calculated with Kurucz's POS lines) in the
FUSE and STIS wavelength ranges where in each case the abundance 
of an individual element X in the construction of the generic (Ca, Sc, Ti, V, Cr, Mn, Co) model atom is 
increased by a factor of ten. Values higher that 1 in the flux ratio indicate stronger lines of element X.

E.g\@. the case of Ti, two lines are much stronger than all others, \Ionww{Ti}{4}{1451.739, 1467.343}.
They are not identified in the observation but at the ten times increased abundance they are clearly
visible in the model.  The same is valid for Cr, where \Ionww{Cr}{4}{1332.415} and \Ionww{Cr}{6}{1417.660} 
are the strongest lines in our models (Fig.\,\ref{fig:geninc}), and for Mn and Co as well. 
This allows us to establish upper abundance limits of about 10\,\% solar for Ti, Cr, Mn, and Co 
(cf\@. the beginning of Sect.\,\ref{sect:analysis}).

\subsubsection{Zinc, germanium, and tin}
\label{sect:zngesn}

\begin{figure*}
  \resizebox{\hsize}{!}{\includegraphics{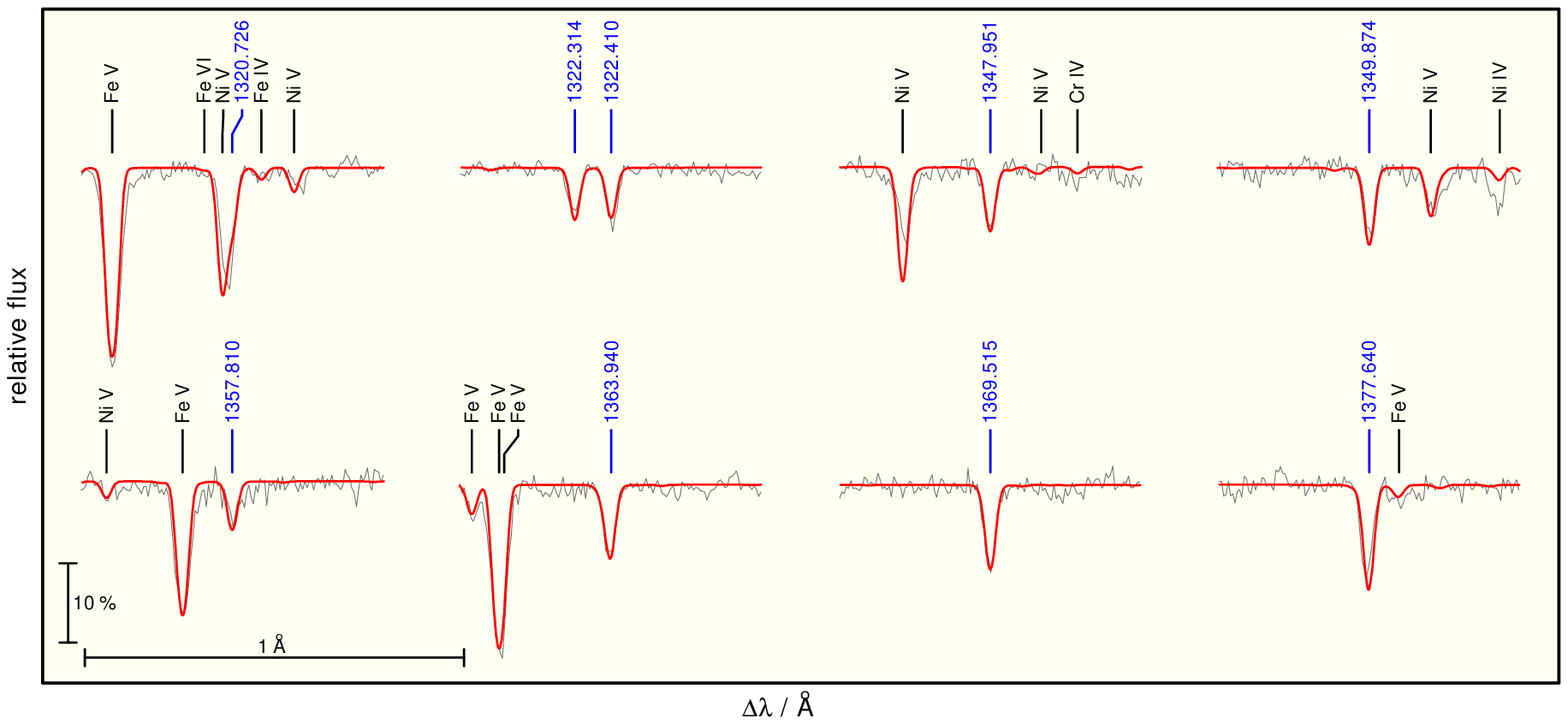}}
  \caption{Theoretical line profiles of the strongest \ion{Zn}{iv} lines in
           the STIS wavelength range compared with the observation.
          } 
  \label{fig:zn}
\end{figure*}

21 \ion{Zn}{iv} lines are newly identified in the STIS observation. These are
almost all that are listed in the NIST\footnote{\url{http://physics.nist.gov/PhysRefData/ASD/lines_form.html}} 
database with relative intensities higher than 100.
Since no individual calculations for \ion{Zn}{iv} transition probabilities are available,
we adapted those of the isoelectronic \ion{Ge}{vi} \citep{rauchetal2012}.
In Fig.\,\ref{fig:zn}, we show nine of them with NIST relative intensities of 200.
All their theoretical line profiles are reproduced at Zn\,$ = -4.89$.

For Ge, we used the same model atom like \citet{rauchetal2012} and determined
Ge\,$ = -5.49$ (Fig.\,\ref{fig:teff}).

We constructed a relatively small Sn model atom.
The only lines for which reliable oscillator strengths are available are
the \ion{Sn}{iii} and \ion{Sn}{iv} resonance lines \citep{morton2000}.
For all other allowed transitions, we follow \citet{werneretal2012} and
set $f = 1$.
We used the \Ionw{Sn}{4}{1314.537} resonance line, like
\citet{vennesetal2005}, to measure the abundance of Sn\,$ = -6.45$.

\subsubsection{Summary of results with chemically homogeneous models}
\label{sect:results}

We can reproduce the entire ultraviolet spectrum of \gb with our chemically homogeneous NLTE models,
with the exception of the \ion{O}{iv} / \ion{}{vi} lines which are obviously affected by
O stratification effects. Current diffusion models yield poor fits to the metal lines \citep{dreizlerwolff1999}.
\Teffw{60\,000 \pm 2000} and 
\loggw{7.60 \pm 0.05} were determined within small error limits.
They are in agreement with 
\citet[\Teffw{60\,920 \pm 993},
       \loggw{7.55 \pm 0.05}]{gianninasetal2011}.
We do not encounter problems in modeling \ion{H}{i} Lyman and Balmer lines simultaneously with
the same \Teff and  \logg like found by \citet[][see Table\,\ref{tab:previous}]{barstowetal2001}.

\begin{figure}
  \resizebox{\hsize}{!}{\includegraphics{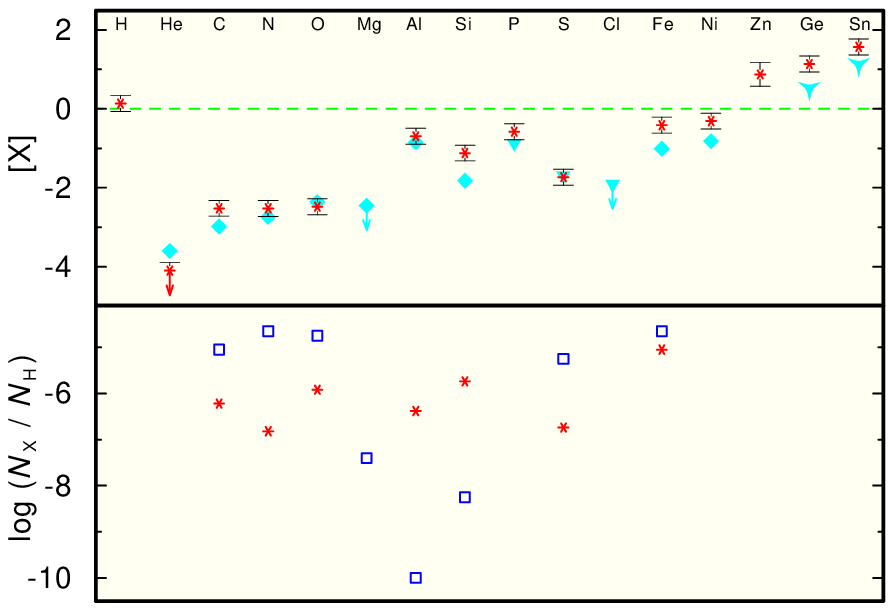}}
  \caption{Top: Photospheric abundances of \gb\ (red stars) compared with solar values
             \citep{asplundetal2009}.
             [X] denotes log ( mass fraction / solar mass fraction ) of element X.
             The dashed, green line shows the solar ratio.
             The arrows indicate upper limits.
             The cyan 
             diamonds \citep{holbergetal2003},
             triangles \citep{vennesetal1996}, and
             tridents \citep{vennesetal2005} are previously determined values.
             Bottom: Comparison of our abundance number ratios (red stars) with predictions of
             diffusion calculations for 
             DA-type (blue squares) WDs \citep{chayeretal1995} with \Teffw{60\,000} and \loggw{7.5}. 
          } 
  \label{fig:chayerDA}
\end{figure}

We can determine all abundances with error limits of 0.2\,dex. In case of Zn, where we adopt
\ion{Ge}{vi} f-values, we estimate that the error is 0.3\,dex.
Our C, N, O, Al, Si, Fe, and Ni abundances (Fig.\,\ref{fig:chayerDA}) agree, 
within error limits, with those of \citet{vennesetal1996,holbergetal2003,vennesetal2005}.
Our values are in general slightly higher. One reason may be the about 6000\,K 
higher \Teff of our final model. The stellar parameters are summarized in Table.\,\ref{tab:results}.

The abundances of all elements but Fe predicted by \citet{chayeretal1995} for a DA-type WD
differ strongly from those that we determined (Fig.\,\ref{fig:chayerDA}).

\subsection{Test of the diffusion impact}
\label{sect:profile}

\begin{figure}
  \resizebox{\hsize}{!}{\includegraphics{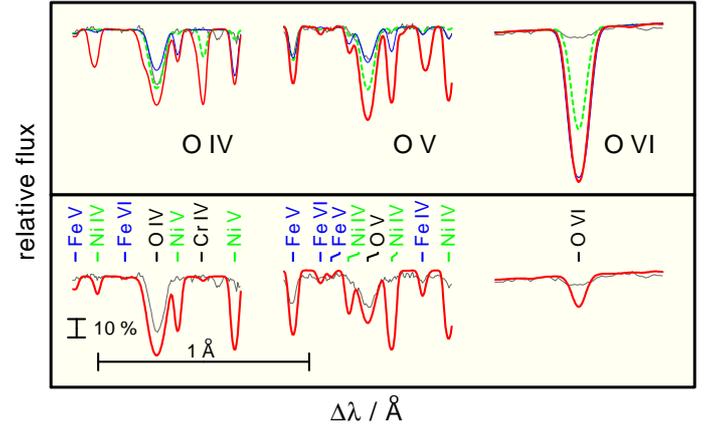}}
  \caption{Same as Fig.\,\ref{fig:ostat}.
           Top: dashed, green TMAP (= chemically homogeneous) model with O\,$ = -4.72$;
                   thin, blue TMAP model with abundance profiles from \citet{dreizlerwolff1999},
                   thick, red NGRT (= diffusion) model.
           Bottom: thick, red NGRT model with an artificially reduced O abundance in the outer atmosphere.
           Note that in the NGRT models, the line strengths of the generic iron-group element (see text) are
           overestimated.
          } 
  \label{fig:odiff}
\end{figure}

In a first step, we simply applied the abundance profiles provided by \citet{dreizlerwolff1999}
to the occupation numbers of He, C, N, O, Si, and Ni in our final model. Figure\,\ref{fig:odiff}
(top panel) shows that this gives a good agreement with \ion{O}{v} while \ion{O}{iv} is now too weak.
The \ion{O}{vi} lines appears even stronger, strengthening the discrepancy. Since the atmospheric
structure was kept fixed in this test, we expected that, if at all, only a self-consistent diffusion model
is able to reproduce the observed \ion{O}{iv} - \ion{}{vi} lines simultaneously.

\subsection{A self-consistent diffusion model}
\label{sect:diffusion}

We used the NGRT\footnote{New generation radiative transport} code \citep{dreizlerwolff1999, schuhetal2002} 
to calculate diffusion models with the same element composition and model atoms like our homogeneous TMAP models. 
The first model shows a strongly increased abundance of the generic model atom that combines Ca, Sc, Ti, V, Cr, Mn, and Co
(Sect\@.\ref{sect:analysis}) and, hence, much too strong lines of the considered elements. The reason is that the 
IrOnIc code \citep{rauchdeetjen2003} calculates a mean atomic weight for the generic atom following

\begin{equation}
A_\mathrm{IG} = \frac{\sum_\mathrm{i=1}^\mathrm{n} r_\mathrm{i} \cdot A_\mathrm{i}}{\sum_\mathrm{i=1}^\mathrm{n} r_\mathrm{i}} ,
\end{equation}

\noindent
where $r_\mathrm{i}$ is the relative mass-fraction (with respect to $r_\mathrm{1} = 1$) and $A_\mathrm{i}$ the atomic
weight of element $i$. The artificially increased number of lines of a single generic element strongly increases its
radiative levitation. Flux blocking by the generic element then leads to stronger gravitational settling of
other elements,
e.g\@. Sn had an abundance below $10^{-17}$ throughout the model atmosphere. The other elements showed
abundances that were partly more than one dex below those of our homogeneous model.
Since we did not want to neglect all opacities of the generic atom, we changed its atomic weight to

\begin{equation}
A_\mathrm{IG} = \sum_\mathrm{i=1}^\mathrm{n} r_\mathrm{i} \cdot A_\mathrm{i} .
\end{equation}

\begin{figure}
  \resizebox{\hsize}{!}{\includegraphics{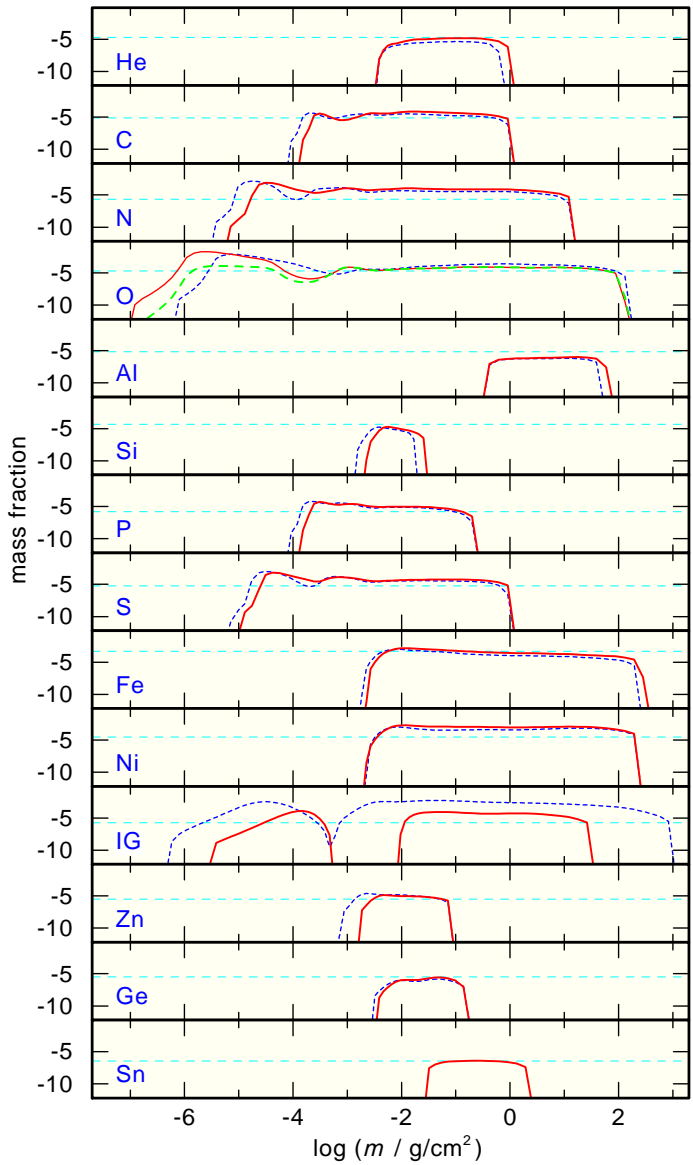}}
  \caption{Abundance profiles in our diffusion model (\Teffw{60\,000}, \loggw{7.60}).
           Short-dashed (blue) lines: unrealistically high abundance of the generic iron-group element (IG, see text),
           thick (red) lines: reduced IG abundance,  
           horizontal long-dashed, thin (cyan) lines: the abundances in our final homogeneous model.
           In the O panel, the thick dashed (green) line shows our modified O-abundance profile
           (see text).
          } 
  \label{fig:depabund}
\end{figure}

\begin{figure}
  \resizebox{\hsize}{!}{\includegraphics{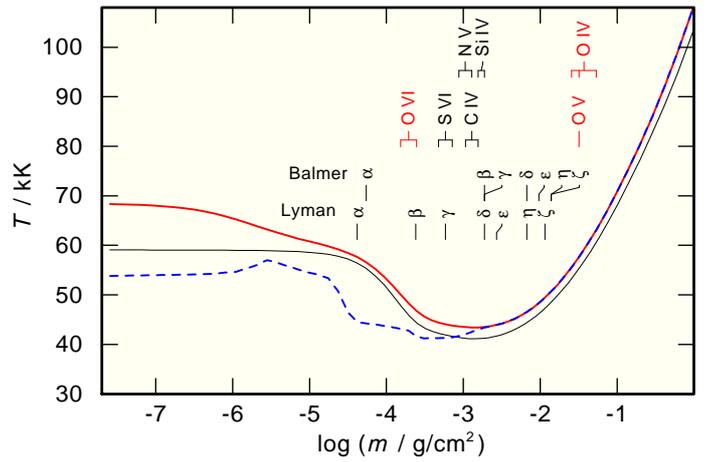}}
  \caption{Temperature structures (\Teffw{60\,000}, \loggw{7.60}) of a pure H model (thin, black),
           our final homogeneous model (thick, red), and a diffusion model (dashed, blue). 
           The formation
           depths of the line cores of the lowest members of the \ion{H}{i} Lyman and Balmer
           series, the \ion{C}{iv}, \ion{N}{v}, \ion{O}{vi}, \ion{Si}{iv}, and \ion{S}{vi}
           resonance doublets, and our strategic  
           \ion{O}{iv} $\lambda\lambda\, 1338.634, 1343.022, 1343.526\,\mathrm{\AA}$
           and 
           \ion{O}{v} $\lambda\, 1371.296\,\mathrm{\AA}$ lines are marked.
          } 
  \label{fig:Tstruct}
\end{figure}

\begin{figure}
  \resizebox{\hsize}{!}{\includegraphics{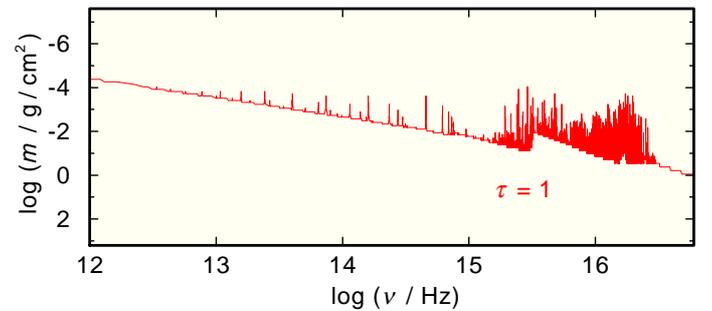}}
  \caption{Optical depth $\tau = 1$ in our final homogeneous model.
          } 
  \label{fig:lp}
\end{figure}

Now, the stratified NGRT models yields depth dependent abundances
(Fig.\,\ref{fig:depabund}) that are closer to those of our homogeneous model, especially Sn appears at
a realistic value. In case of He, C, N, O, Si, and Ni the abundance profiles are similar to those of 
\citep{dreizlerwolff1999}. The changed atmospheric structure is shown in Fig.\,\ref{fig:Tstruct}.
It is interesting to note that most of the lines and all continua are formed at
$\log m\, \sga\, -3$ (Fig.\,\ref{fig:lp}) while deviations in the temperature structure are noticeable only
outside of this region.
The resulting spectrum (Fig.\,\ref{fig:TMAPNGRT}) of the stratified model is, compared with the
homogeneous model, no improvement. While \ion{Fe}{v} lines match the observation at
about \Teffw{55\,000}, it can be extrapolated that \ion{Fe}{iv} lines are much too strong for 
\Teff $\sla\, 70\,000$\,K. \ion{Ni}{iv} and \ion{Ni}{v} lines are much too strong because the
Ni abundance is enhanced (Fig.\,\ref{fig:depabund}) in the line-forming regions and can,
thus, not be used for a \Teff estimate.

\onlfig{
\begin{figure*}
  \resizebox{\hsize}{!}{\includegraphics{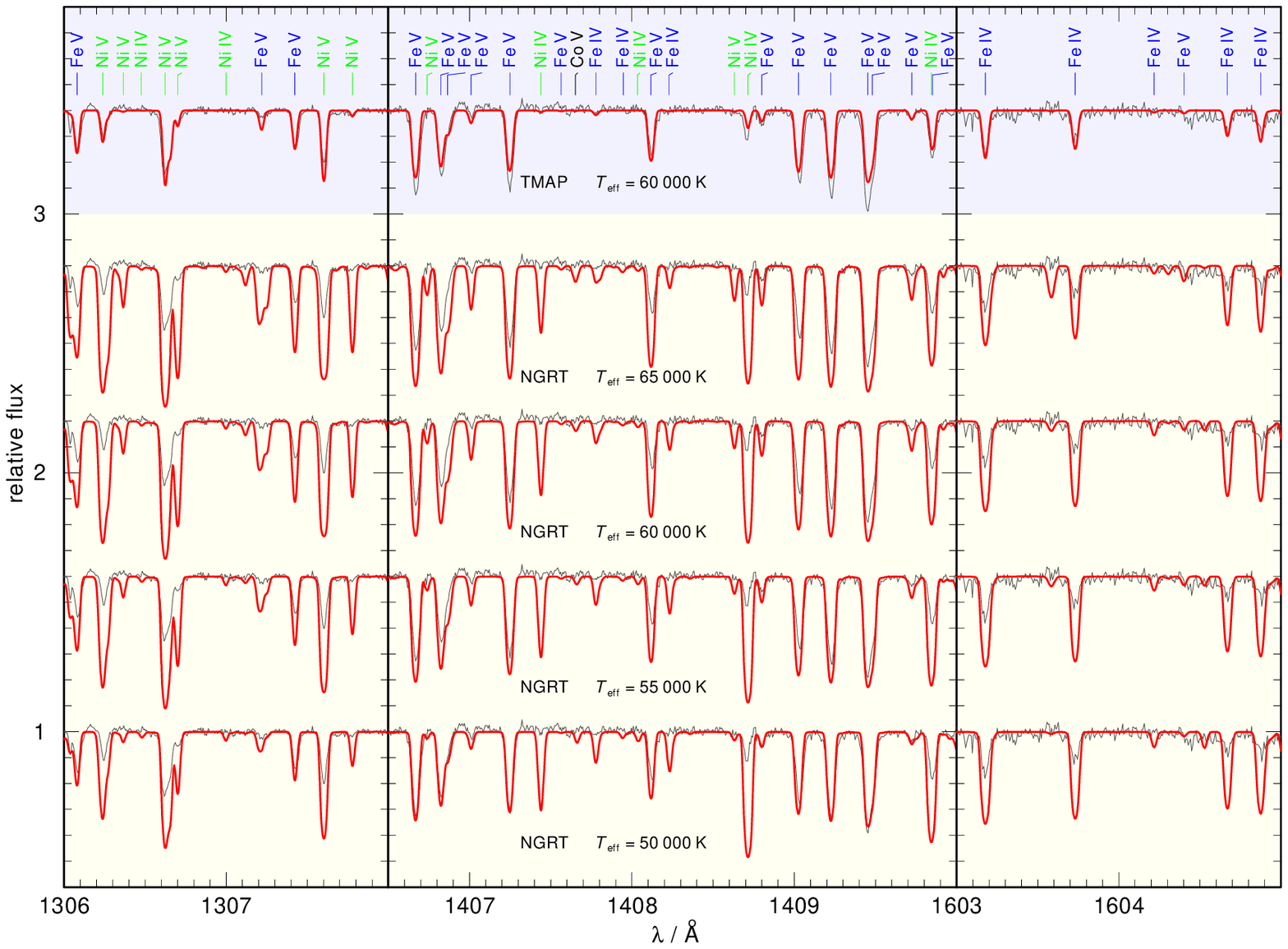}}
  \caption{Comparison of our model spectra (all calculated with \loggw{7.60}) with the STIS observation.
           Top: Final homogeneous model (TMAP).
           The stratified NGRT models have \Teff = 65\,000\,K, 60\,000\,K, 55\,000\,K, and 50\,000\,K 
           (from top to bottom). The lines in the models are marked.
          } 
  \label{fig:TMAPNGRT}
\end{figure*}
}

In the stratified models, the  \ion{O}{iv} and \ion{O}{v} lines
are now mucher stronger than observed and \ion{O}{vi} appears at the same strength that resulted
from our diffusion test (Sect.\,\ref{sect:profile}). The O abundance profile (Fig.\,\ref{fig:depabund})
shows a strong increase for $\log m < -4$. Only by the introduction of an artificial abundance reduction by a factor
of $m / 1585$ for $\log m\, \sla -3.2$, we achieve an acceptable agreement of \ion{O}{v} and \ion{O}{vi}
(Fig.\,\ref{fig:odiff}). \ion{O}{iv} is still slightly too strong because the abundance and, thus, the lines 
(including a blend at \ion{O}{iv}) of the generic iron-group atom (Sect.\,\ref{sect:models}) are overestimated 
by the NGRT model (Fig.\,\ref{fig:depabund}). 
Based on this numerical exercise,
it may be speculated that a weak stellar wind or an other, unknown process that is not considered by
NGRT is responsible for the lower oxygen abundances in the outer atmosphere.

We can conclude two things. A generic model atom is obviously not suited for a diffusion calculation 
due to the strongly enhanced number of lines for a single atom in the modeling process. The NGRT diffusion models 
yield partly too low abundances in the line-forming regions
and, thus, cannot reproduce the metal line properly. An additional, weak wind may
be necessary to increase the metal abundances in the line-forming regions.

\subsection{The extreme-ultraviolet spectrum}
\label{sect:euv}

The inability to model the EUVE spectrum with chemically homogeneous
atmospheres \citep{holbergetal1989} was the reason to investigate stratified photospheres 
\citep[e.g\@.][]{koester1991}. \citet{lanzetal1996} demonstrated, that it is possible to consistently match
the optical, UV, and EUV data with homogeneous NLTE models with the same \Teff and chemical composition.

\begin{figure}
  \resizebox{\hsize}{!}{\includegraphics{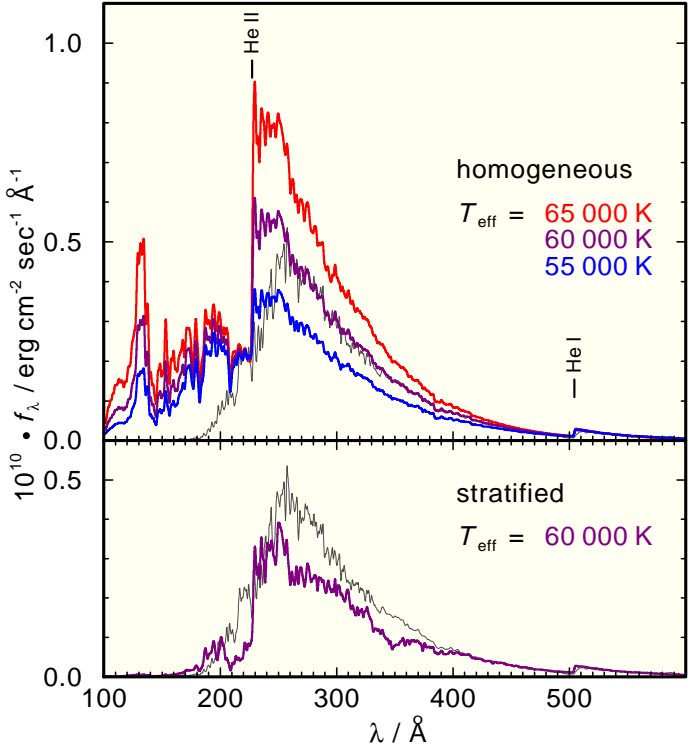}}
  \caption{Top: Comparison of three synthetic spectra (\Teff = 55\,000, 60\,000, and 65\,000\,K)  of 
                chemically homogeneous models with the EUVE observation.
           The wavelengths of ground-state absorption thresholds of He ions are indicated. 
           Bottom: Comparison of a stratified model \Teffw{60\,000} with the observation.
          } 
  \label{fig:euv}
\end{figure}

We calculated EUV spectra from our model grid with 193\,584 frequency points within 100\,\AA\ $\le \lambda \le$ 930\,\AA,
and Kurucz's LIN line lists \citep[theoretical and laboratory measured lines, in total 8\,135\,405 lines of Ca - Ni 
in our wavelength interval,][]{kurucz2009}. 
These spectra were processed with the recently registered VO tool TEUV\footnote{\url{http://astro.uni-tuebingen.de/~TEUV}}
that corrects synthetic stellar fluxes for interstellar absorption below 911\,\AA. It simulates radiative bound-free 
absorption of the lowest ionization states of H, He, C, N, and O using Opacity Project data \citep{seatonetal1994}.
Two interstellar components with different radial and turbulent velocities, temperatures, and column densities can 
be considered. 
Figure\,\ref{fig:euv} shows the comparison of synthetic and observed EUV spectra.
Our synthetic spectra were normalized to the measured FUSE flux of $1.347 \times 10^{-11} \mathrm{erg\,cm^{-2}\,s^{-1}\,\AA^{-1}}$ 
at 920\,\AA. Then, the interstellar column densities are adjusted, to match the EUVE flux 
$N_\ion{H}{i}$   for 530\,\AA,
$N_\ion{He}{i}$  for 470\,\AA, and
$N_\ion{He}{ii}$ for 220\,\AA. Since our models do not reproduce the measured flux between 250\,\AA\, and
the \ion{He}{ii} ground state threshold, $N_\ion{He}{ii}$ is not reliable.
Table\,\ref{tab:ismcol} shows the applied $N_\ion{H}{i}$ and $N_\ion{He}{i}$ values compared with
the literature values. Our $N_\ion{H}{i}$ values, necessary to match the EUVE flux level, 
are about a factor of two higher than $\log\ (N_\ion{H}{i}\,/\,\mathrm{cm^2}) = 18.34^{+0.08}_{-0.10}$ 
that we determined previously from \ion{H}{i} Lyman-line fits (Sect.\,\ref{sect:ism}).
$\log N_\ion{N}{i} = 13.87$ and $\log N_\ion{O}{i} = 14.86$ were adopted from \citet{lemoineetal2002}.

\begin{table}[]\centering
\caption{Logarithmic ISM column densities for homogeneous models with different \Teff to match the EUVE flux level of \gb.}
\label{tab:ismcol}
\begin{tabular}{rr@{.}lr@{.}ll}
\hline
\Teff / K & \multicolumn{2}{c}{$N_\ion{H}{i}$} & \multicolumn{2}{c}{$N_\ion{He}{i}$} \\
\hline
\multicolumn{5}{l}{homogeneous (TMAP)} & \\
\hbox{}\hspace{3mm}55\,000 & 18&53 & 17&45 & \\
                   60\,000 & 18&59 & 17&45 & \\
                   65\,000 & 18&64 & 17&45 & \\
\hline
\multicolumn{5}{l}{stratified (NGRT)} \\
                   60\,000 & 18&60 & 17&45 & \\
\hline
\multicolumn{5}{l}{literature} & \\
                   59\,250 & 18&23 & 17&16 & \citet[][HUT]{kimbleetal1993}     \\
                   54\,000 & 18&27 & 17&16 & \citet[][EUVE]{dupuisetal1995}    \\
                   55\,200 & 18&32 & 17&26 & \citet[][EUVE]{lanzetal1996}      \\
                   56\,000 & 18&32 & 17&15 & \citet[][EUVE]{dreizlerwolff1999} \\
                   53\,000 & 18&28 & 17&16 & \citet[][EUVE]{venneslanz2001}    \\
                   54\,000 & 18&33 & 17&34 & \citet[][J-PEX]{cruddaceetal2002} \\
\hline
\end{tabular}
\end{table}

The overall agreement of our homogeneous models with \Teffw{60\,000} at wavelengths $\lambda \sga$\,250\,\AA\ is very good, 
especially the interval 360\,\AA\,$\sla \lambda \sla$\,450\,\AA\ is excellently matched in detail. 
Models with \Teffw{65\,000} and \Teffw{55\,000} yield much too high and too low fluxes, respectively. At 
$\lambda \sla$\,250\,\AA\ the theoretical flux is too high in all models, even at \Teffw{55\,000}. 
A stratified model (Fig.\,\ref{fig:euv}) with \Teffw{60\,000}
fails to reproduce the flux between 250\,\AA\,$\sla \lambda \sla$\,420\,\AA\ and has a too-high flux
at $\lambda \sla$\,200\,\AA.

Both, our homogeneous and our stratified models, fail to reproduce the entire EUV spectrum of \gb.
It seems likely that there may be some stratification in the atmosphere but we don't yet know how to
distribute the various atomic species with depth. This is a challenge for theorists.

\subsection{Mass and distance}
\label{sect:distance}

\begin{figure}
  \resizebox{\hsize}{!}{\includegraphics{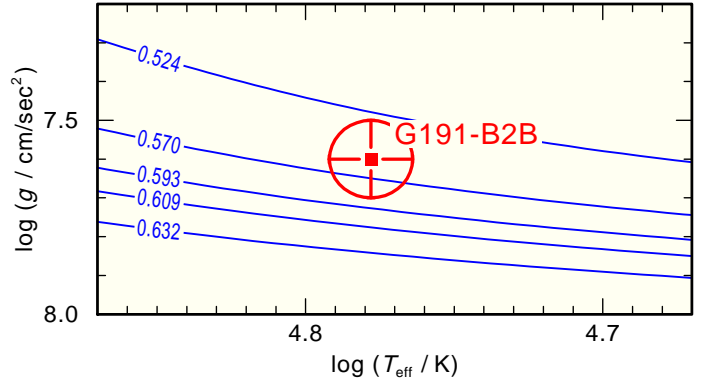}}
  \caption{Location of \gb in the $\log$ \Teff - \logg plane (the ellipse indicates the errors of our analysis) 
           compared with evolutionary tracks
           for hydrogen-rich white dwarfs \citep{renedoetal2010} labeled with the respective
           stellar masses (in \Msol).} 
  \label{fig:evo}
\end{figure}

A stellar mass of $M = 0.555^{+0.035}_{-0.029} M_\odot$ and
a luminosity of $\log (L / L_\odot) = 0.63^{+0.37}_{-0.34}$
are determined by comparison with evolutionary models (Fig.\,\ref{fig:evo})
for old white dwarfs (metallicity $z = 0.001$).

\begin{table}\centering
  \caption{Parameters of \gb as derived by our analysis.}         
\label{tab:results}
\setlength{\tabcolsep}{.4em}
\begin{tabular}{rlr@{.}lr@{.}lr@{.}l}
\hline
\hline
\noalign{\smallskip}                                                                                                     
\multicolumn{2}{l}{$T_\mathrm{eff}\,/\,$K}              & \multicolumn{2}{l}{$60\,000\pm 2000$}             & \multicolumn{4}{l}{}    \\
\multicolumn{2}{l}{$\log\ ( g\,/\,\mathrm{cm/sec^2} )$} & \multicolumn{2}{l}{$7.60\pm 0.05$}                & \multicolumn{4}{l}{}    \\
\hline
\noalign{\smallskip}                                                                                          
&                          & \multicolumn{2}{c}{mass}   & \multicolumn{2}{c}{number}  & \multicolumn{2}{c}{}                    \\
\cline{3-6}                     
\multicolumn{8}{c}{}                                                                                             \vspace{-5mm}\\
& element                  & \multicolumn{2}{c}{}       & \multicolumn{2}{c}{}        & \multicolumn{2}{c}{~~~~~[X]} \vspace{-2mm}\\
&                          & \multicolumn{4}{c}{fraction}                           & \multicolumn{2}{c}{}                    \\
\cline{2-8}                     
\noalign{\smallskip}                                                                                   
\smspr & \mmspr H                        & $ 9$&$99\times 10^{-1}$ & $ 1$&$0              $ & $  0$&$132$ \\
       & \mmspr He                       & $<1$&$98\times 10^{-5}$ & $<5$&$0\times 10^{-6}$ & $<-4$&$099$ \\
       & \mmspr C                        & $ 7$&$15\times 10^{-6}$ & $ 6$&$0\times 10^{-7}$ & $ -2$&$520$ \\
       & \mmspr N                        & $ 2$&$08\times 10^{-6}$ & $ 1$&$5\times 10^{-7}$ & $ -2$&$522$ \\
       & \mmspr O                        & $ 1$&$90\times 10^{-5}$ & $ 1$&$2\times 10^{-6}$ & $ -2$&$479$ \\
       & \mmspr Al                       & $ 1$&$12\times 10^{-5}$ & $ 4$&$2\times 10^{-7}$ & $ -0$&$695$ \\
       & \mmspr Si                       & $ 5$&$01\times 10^{-5}$ & $ 1$&$8\times 10^{-6}$ & $ -1$&$123$ \\
       & \mmspr P                        & $ 1$&$54\times 10^{-6}$ & $ 5$&$0\times 10^{-8}$ & $ -0$&$579$ \\
       & \mmspr S                        & $ 5$&$72\times 10^{-6}$ & $ 1$&$8\times 10^{-7}$ & $ -1$&$733$ \\
       & \mmspr Ti                       & $<3$&$13\times 10^{-7}$ & $<1$&$1\times 10^{-7}$ & $<-0$&$100$ \\
       & \mmspr Cr                       & $<1$&$66\times 10^{-6}$ & $<5$&$1\times 10^{-7}$ & $<-0$&$100$ \\
       & \mmspr Mn                       & $<1$&$08\times 10^{-6}$ & $<3$&$2\times 10^{-7}$ & $<-0$&$100$ \\
       & \mmspr Fe                       & $ 4$&$98\times 10^{-4}$ & $ 9$&$0\times 10^{-6}$ & $ -0$&$414$ \\
       & \mmspr Co                       & $<4$&$92\times 10^{-7}$ & $<1$&$2\times 10^{-7}$ & $<-0$&$100$ \\
       & \mmspr Ni                       & $ 3$&$49\times 10^{-5}$ & $ 6$&$0\times 10^{-7}$ & $ -0$&$310$ \\
       & \mmspr Zn                       & $ 1$&$30\times 10^{-5}$ & $ 2$&$0\times 10^{-7}$ & $  0$&$873$ \\
       & \mmspr Ge                       & $ 3$&$24\times 10^{-6}$ & $ 4$&$5\times 10^{-8}$ & $  1$&$135$ \\
       & \mmspr Sn                       & $ 3$&$53\times 10^{-7}$ & $ 3$&$0\times 10^{-9}$ & $  1$&$569$ \\
\hline
\noalign{\smallskip}                                                                                          
\multicolumn{2}{l}{\ebv}                                & \multicolumn{2}{l}{$0.0005 \pm 0.0005$}           & \multicolumn{4}{l}{}    \\
\multicolumn{2}{l}{$\log\ ( \nh\,/\,\mathrm{cm^2} )$}   & \multicolumn{2}{l}{$18.34^{+0.08}_{-0.10}$}       & \multicolumn{4}{l}{}    \\
\noalign{\smallskip}                                                                                          
\multicolumn{2}{l}{$\log\ ( \deh\,/\,\mathrm{cm^2} )$}  & \multicolumn{2}{l}{$13.54^{+0.05}_{-0.06}$}       & \multicolumn{4}{l}{}    \\
\multicolumn{2}{l}{$d\,/\,\mathrm{pc}$}                 & \multicolumn{2}{l}{$62 \pm 4$}                    & \multicolumn{4}{l}{}    \\
\multicolumn{2}{l}{$M\,/\,M_\odot$}                     & \multicolumn{2}{l}{$0.555 ^{+0.035}_{-0.029}$}    & \multicolumn{4}{l}{}    \\
\noalign{\smallskip}                                                                                              
\multicolumn{2}{l}{$R\,/\,R_\odot$}                     & \multicolumn{2}{l}{$0.0195 ^{+0.0004}_{-0.0005}$} & \multicolumn{4}{l}{}    \\
\noalign{\smallskip}                                                                                              
\multicolumn{2}{l}{$\log\ ( L\,/\,L_\odot )$}           & \multicolumn{2}{l}{$0.63^{+0.37}_{-0.34}$}        & \multicolumn{4}{l}{}    \\
\noalign{\smallskip}
\hline         
\hline
\end{tabular}
\end{table}

We calculated the spectroscopic distance following the flux calibration 
of \citet{heberetal1984a} for $\lambda_\mathrm{eff} = 5454\,\mathrm{\AA}$,

\begin{equation}
d[\mathrm{pc}]=7.11 \times 10^{-4} \cdot \sqrt{H_\nu\cdot M \times 10^{0.4\, m_{\mathrm{v}_0}-\log g}} \,\, ,$$
\end{equation}

\noindent
with $m_\mathrm{V_o} = m_\mathrm{V} - 2.175 c$, $c = 1.47 E_\mathrm{B-V}$, and 
the Eddington flux $H_\nu = 1.109 \times 10^{-3}\, \mathrm{erg\, cm^{-2}\, s^{-1}\, Hz^{-1}}$ at $\lambda_\mathrm{eff}$ of our final model atmosphere.
We used 
$E_\mathrm{B-V}=0.0005 \pm 0.0005$ (Sect.\,\ref{sect:ism}), 
$M = 0.555^{+0.035}_{-0.029} M_\odot$, and
$m_{\mathrm{V}} = 11.7228 \pm 0.0082$ \citep{vanleeuwen2007},
and derived a distance of 
$d=62 \pm 4$\,pc 
and a height above the 
Galactic plane of 
$z=8 \pm 1$\,pc. 
This is in agreement with 
the HIPPARCOS\footnote{\url{http://www.rssd.esa.int/index.php?project=HIPPARCOS}} parallax measurement 
\citep[\object{HIP23692}]{vanleeuwen2007} of $d=59.88^{\,\,\,+9.05}_{-12.95}$\,pc
and the XHIP\footnote{Extended HIPPARCOS compilation} value of $d=57.96 \pm 10.31$\,pc \citep{andersonfrancis2012}.
The spectroscopic distance of $d=55.84 \pm 0.86$\,pc determined by \citet{holbergetal2008} is slightly smaller,
this error estimate, however, appears to be too optimistic.

\section{TheoSSA: synthetic stellar spectra on demand}
\label{sect:theossa}

\begin{figure}
  \resizebox{\hsize}{!}{\includegraphics{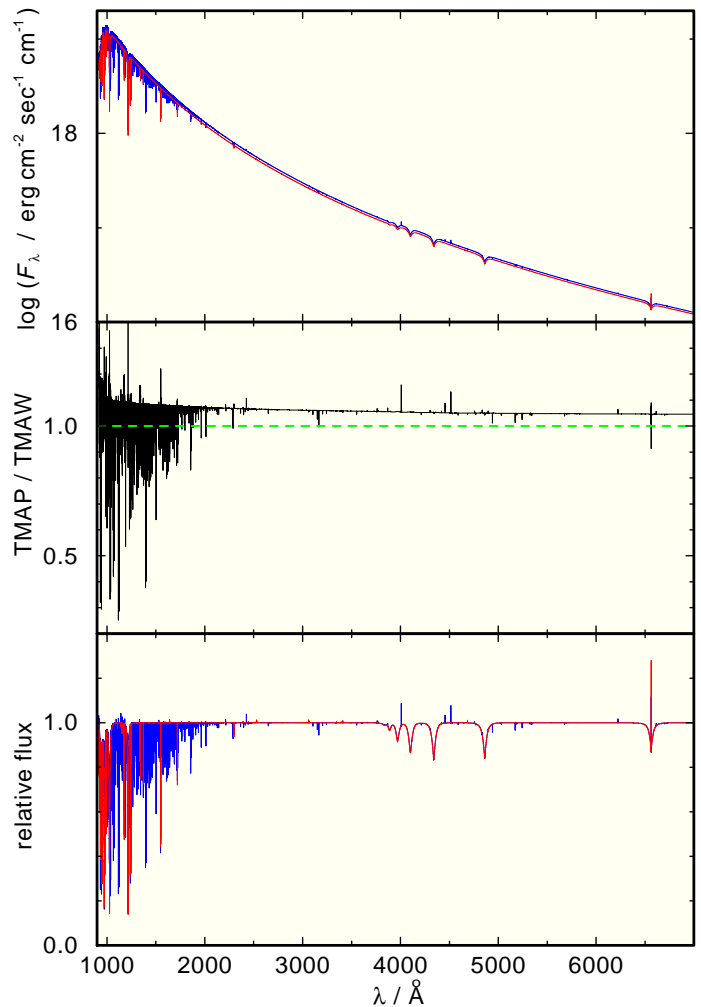}}
  \caption{Comparison of our final TMAP model (blue, thick) with a
           TMAW model (red, thin).
           Top: astrophysical fluxes at the stellar surface,
           middle: ratio TMAP/TMAW flux,
           bottom: normalized fluxes.
          } 
  \label{fig:tmawtmap}
\end{figure}

At the end, we want to compare our final TMAP model flux with an SED, that was calculated with the
TMAW tool (Sect.\,\ref{sect:intro}) and considers H, He, C, N, and O only.
Figure\,\ref{fig:tmawtmap} shows that the TMAP flux is higher everywhere at $\lambda > 911$\,\AA.
The reason is strong metal-line blanketing at $\lambda < 911$\,\AA\ that causes a flux increase
at longer wavelengths. It amounts to about 10\,\% at 1000\,\AA\ and to about 5\,\% at 7000\,\AA.
The lower panel of Fig.\,\ref{fig:tmawtmap} illustrates that the theoretical line profiles of
the \ion{H}{i} Balmer series are almost identical in both, TMAP and TMAW models, with the
exception of an increased emission reversal in the line core of H\,$\alpha$ in the TMAP model.

The TheoSSA database contains currently TMAP SEDs of a dozen standard stars.
Some of them are represented by different models for comparison, because they
were initially calculated for the calibration \citep{vernetetal2008a, vernetetal2008b, vernetetal2010} 
of ESO's\footnote{European Southern Observatory} second generation VLT\footnote{Very Large Telescope}
instrument XSHOOTER\footnote{\url{http://www.eso.org/sci/facilities/paranal/instruments/xshooter}}
\citep{vernetetal2011} while the parameters
of \citet{gianninasetal2011} and \citet{giammicheleetal2012} were published later (Table\,\ref{tab:theocontent}).

\begin{table}[ht!]\centering
\caption{Standard star SEDs (references for \Teff and \logg are given) presently available in TheoSSA. 
         WD numbers are from \citet{mccooksion1999}.}
\label{tab:theocontent}
\setlength{\tabcolsep}{.4em}
\begin{tabular}{llccr@{.}l}
\hline
\noalign{\smallskip}
&&&
$T_\mathrm{eff}$ & 
\multicolumn{2}{c}{$\log g$} \vspace{-2mm} \\
name & WD no. & spectral type &
& 
\multicolumn{2}{c}{}\vspace{-2mm} \\
&&&
[K] & 
\multicolumn{2}{c}{[$\mathrm{cm/sec^2}$]} \\
\noalign{\smallskip}
\hline
\noalign{\smallskip}
\object{EG\,274}    & 1620$-$391 & DA2 (+ G5V) & 24\,276 & \hbox{}\hspace{3mm}8&01\tablefootmark{a} \\ 
                    &            &             & 25\,980 &                    7&96\tablefootmark{b, c} \\
                    
\object{Feige\,67}  &            & Op+WDsd     & 75\,000 &                    5&20\tablefootmark{d} \\
                    
\object{Feige\,110} & 2317$-$054 & sdO         & 40\,000 &                    5&00\tablefootmark{e} \\
                    
\gb                 & 0501$+$527 & DA0         & 58\,883 &                    7&46\tablefootmark{a} \\ 
                    &            &             & 61\,193 &                    7&49\tablefootmark{f} \\
                    &            &             & 60\,920 &                    7&55\tablefootmark{b} \\
                    &            &             & 60\,000 &                    7&55\tablefootmark{g} \\
                    
\object{G\,93$-$48} & 2149$+$021 & DAZ3        & 18\,100 &                    7&85\tablefootmark{h} \\
                    &            &             & 18\,170 &                    8&01\tablefootmark{b} \\
                    
\object{GD\,50}     & 0346$-$011 & DA2         & 40\,550 &                    9&22\tablefootmark{i} \\
                    &            &             & 42\,700 &                    9&20\tablefootmark{b} \\
                    
\object{GD\,71}     & 0549$+$158 & DA1         & 32\,747 &                    7&68\tablefootmark{f} \\
                    &            &             & 32\,780 &                    7&83\tablefootmark{i} \\
                    &            &             & 33\,590 &                    7&93\tablefootmark{b} \\
                    
\object{GD\,108}    & 0958$-$073 & sdB         & 22\,908 &                    5&30\tablefootmark{j} \\
                    
\object{GD\,153}    & 1254$+$223 & DA1.5       & 38\,205 &                    7&89\tablefootmark{i} \\
                    &            &             & 38\,686 &                    7&66\tablefootmark{f} \\
                    &            &             & 40\,590 &                    7&93\tablefootmark{b} \\
                    
\object{HZ\,2}      & 0410$+$117 & DA3         & 20\,600 &                    7&90\tablefootmark{h} \\
                    &            &             & 21\,600 &                    7&98\tablefootmark{b} \\
                    
\object{HZ\,43A}    & 1314$+$293 & DA1+dM3e    & 51\,116 &                    7&90\tablefootmark{k} \\
                    &            &             & 56\,800 &                    7&89\tablefootmark{b} \\
                    
\object{Sirius\,B}  & 0642$-$166 & DA2         & 24\,826 &                    8&60\tablefootmark{k} \\
                    &            &             & 25\,970 &                    8&57\tablefootmark{b, c} \\
\hline
\end{tabular} 
\tablefoot{~\\
\tablefoottext{a}{assumed}
\tablefoottext{b}{\citet{gianninasetal2011}}
\tablefoottext{c}{\citet{giammicheleetal2012}}
\tablefoottext{d}{\citet{bauerhusfeld1995}}
\tablefoottext{e}{\citet{heberetal1984b}, He mass fraction of 0.107}
\tablefoottext{f}{\citet{finleyetal1997}}
\tablefoottext{g}{this work}
\tablefoottext{h}{\citet{guseinovetal1983}}
\tablefoottext{i}{\citet{barstowetal2001}}
\tablefoottext{j}{\citet{kilkennyetal1988}}
\tablefoottext{k}{\citet{beuermannetal2006}}
}
\end{table}

\section{Accuracy of flux calibration with \gb}
\label{sect:fluxcal}

\begin{figure}
  \resizebox{\hsize}{!}{\includegraphics{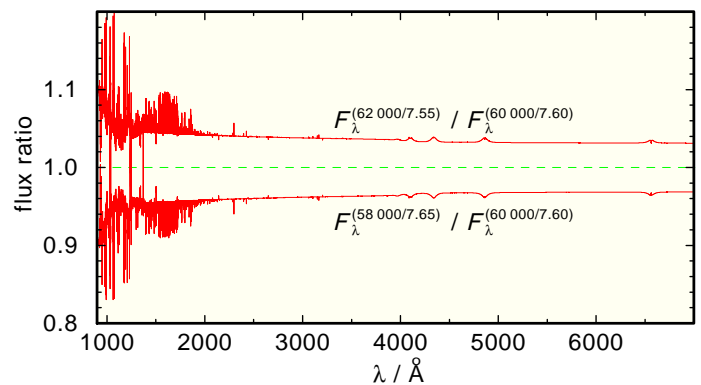}}
  \caption{Flux ratio of two TMAP models at the edge of the \Teff and \logg
           error ranges and our final model.
          } 
  \label{fig:fluxcal}
\end{figure}

Our spectral analysis was performed using state-of-the-art atomic data and
model-atmosphere code. The best reproduction of UV and optical spectra was 
achieved with chemically homogeneous models. In Fig.\,\ref{fig:fluxcal},
we compared two models at the edge of our error ranges in \Teff and \logg.
The deviation in the continuum flux of two TMAP model SEDs is $\approx 3$\,\% 
in the optical and $\approx 5$\,\% in the FUV. A systematic error is
present due to the uncertainty of the used atomic data, such as oscillator
strengths where it is typically $\approx 15$\,\% for a single line.
The employment of many lines of many ions of many atoms in a spectral analysis 
minimizes the propagation of these uncertainties  into the errors of the
main photospheric properties like \Teff, \logg, and the abundances.
An additional systematic error may be present between individual model-atmosphere
packages \citep[e.g\@.][]{rauch2008} because of differences in coding,
approximations, etc.

The situation for the flux calibration is, however, not that strongly dependent
on the exact \Teff and \logg values (the latter is even less important).
E.g\@. in the case of \gb and our errors (3\,\% in \Teff and 0.05\,dex in \logg),
a normalization to a precisely measured brightness will reduce the deviation
between model SED and observation much below 1\,\% in the optical and infrared.
The residuals among the three primary stars
\gb, 
\object{GD\,71}, and 
\object{GD\,153} 
are generally sub percent at the longer wavelengths \citep{bohlin2007}
The remaining deviation in the FUV wavelength range is presently less than 2\,\%.
Further improvement is essentially dependent on the reliability of the atomic
data (Sect.\,\ref{sect:conclusions}).

\section{Conclusions}
\label{sect:conclusions}

The TheoSSA service is designed to provide theoretical stellar SEDs of any kind in 
VO-compliant form. Its efficiency is strongly increasing if more different 
model-atmosphere groups provide their SEDs with a proper description in their 
respective meta data. 
The establishment of a database of spectrophotometric standard stars is an 
opportunity to use the same model SEDs for astrophysical flux calibration.
Many model-atmosphere groups have their own best models for some of these stars,
for which a common base for comparison arises. 
Differences in the algorithms for considered physics, assumptions, and
approximations in different model-atmosphere codes, lead to systematic 
deviations in general.

\begin{figure}
  \resizebox{\hsize}{!}{\includegraphics{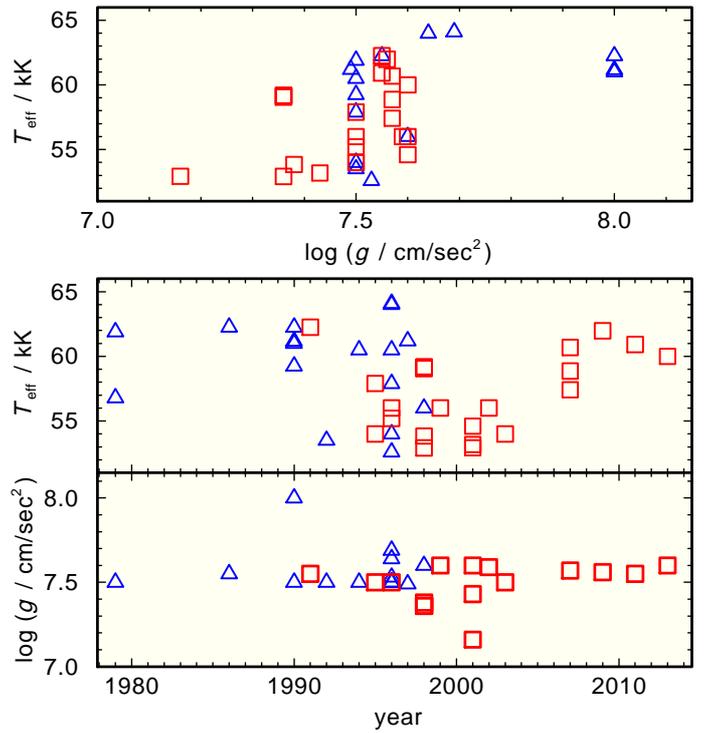}}
  \caption{Determined \Teff and \logg values in the last 34 years.
           Blue triangles denote analyses with LTE models,
           red squares those with NLTE models (Table\,\ref{tab:previous}).           
           The result of \citet[\loggw{5.95}]{koesteretal1979} is outside the top and bottom panels.
          } 
  \label{fig:tefflogghist}
\end{figure}

Figure\,\ref{fig:tefflogghist} shows the temporal development of the
\Teff and \logg determinations of \gb. While both values had a large scatter in the
1990s (error ranges are not shown for clarity), the
three most recent analyses, that are all based on sophisticated NLTE model-atmosphere
techniques, show a good agreement within relatively narrow error ranges of about
3\,\% in \Teff and 0.05\,dex ($\approx 12$\,\%) in \logg. 
Ironically, these latest results agree quite well with the very first
line-profile analysis presented by \citet{holbergetal1986} performed with
a Ly\,$\alpha$ line-profile fit with a simple, pure-H LTE model atmosphere.
The TheoSSA database may help to get closer to the intended goal of 1\,\% 
accuracy in absolute flux calibration.

We presented here our spectral analysis of \gb to demonstrate the current
state-of-the-art. We are presently able to reproduce the observed spectrum
from 250\,\AA\ to the infrared. The EUV part from 150\,\AA\ to 250\,\AA\
cannot be modeled, neither by our homogeneous nor by our stratified models.
The reason is unknown.

A similar analysis of the UV spectrum of the calibration star
\object{BD\,$+28\degr 4211$}
(\Teffw{82\,000 \pm 5000}, \loggw{6.2^{+0.3}_{-0.1}})
was just published by \citet{latouretal2013}.

Model-atmosphere codes have arrived now at a high level of
sophistication, and we already encounter problems getting reliable atomic data to
reproduce the high-resolution and high-S/N spectra that are obtained with
presently available instruments. This is a challenge for atomic physicists
to be prepared for upcoming telescopes and instruments.

\begin{acknowledgements}
TR is supported by the German Aerospace Center (DLR, grant 05\,OR\,1301).  
The GAVO project at T\"ubingen has been supported by the Federal Ministry of Education and
Research (BMBF, grants  05\,AC\,6\,VTB, 05\,AC\,11\,VTB).
AstroGrid-D was funded by the BMBF (01\,AK\,804\,[A-G]).
The bwGRiD\footnote{http://www.bw-grid.de/en/the-bwgrid/} is funded within the framework of the D-Grid Project by the BMBF.
This research has made use of the SIMBAD database, operated at CDS, Strasbourg, France.
This research has made use of NASA's Astrophysics Data System.
This work used the WRPLOT visualization software developed by Wolf-Rainer Hamann (Potsdam) and the WRPLOT team.
Some of the data presented in this paper were obtained from the 
Mikulski Archive for Space Telescopes (MAST). STScI is operated by the 
Association of Universities for Research in Astronomy, Inc., under NASA 
contract NAS5-26555. Support for MAST for non-HST data is provided by 
the NASA Office of Space Science via grant NNX09AF08G and by other 
grants and contracts.
The TEUV tool (\url{http://astro-uni-tuebingen.de/~TEUV}) used to
apply interstellar corrections to theoretical spectra was constructed
as part of the activities of the German Astrophysical Virtual Observatory.
The TIRO service (\url{http://astro-uni-tuebingen.de/~TIRO}) used to calculate
opacities for this paper was constructed as part of the
activities of the German Astrophysical Virtual Observatory.
\end{acknowledgements}

\bibliographystyle{aa}
\bibliography{22336}

\end{document}